\documentclass[10pt,a4paper]{article}
\pdfoutput=1
 \usepackage{jcappub}

\allowdisplaybreaks

\usepackage{ifthen}

\usepackage{amsmath} 
\usepackage{amssymb}
\usepackage{amsthm}
\usepackage{amsfonts}
\usepackage{array}
\usepackage{multirow}
\usepackage{bm}

\usepackage{graphicx}
\usepackage{caption}
\usepackage{subfig}
\usepackage{float}
\usepackage{color}
\usepackage{epstopdf}
\usepackage{booktabs}

\usepackage{appendix}
\usepackage{enumerate}
\usepackage{cleveref}
\usepackage{url}

\newcommand{\fnl}{f_{\rm NL}}

\newcommand{\gnl}{g_{\rm NL}}
\newcommand{\tnl}{\tau_{\rm NL}}
\newcommand{\q}{\mathbf{q}}
\newcommand{\p}{\mathbf{p}}
\newcommand{\x}{\mathbf{x}}
\def\k{\mathbf{k}}
\newcommand{\ps}{\mathbf{\Psi}}
\newcommand{\na}{\mathbf{\nabla}}
\newcommand{\vp}{\varphi}

\newcommand{\MA}{{\rm M}}

\newcommand{\kd}{\kappa_2}

\newcommand{\kt}{\kappa_3}
\newcommand{\ktl}{\kappa_3^{(1)}}

\newcommand{\dkt}{\frac{d\kt/d\MA}{d\ln\sigma^{-1}/d\MA}}
\newcommand{\dktl}{\frac{d\kt^{(1)}/d\MA}{d\ln\sigma^{-1}/d\MA}}

\newcommand{\dk}{\frac{\derivd\k}{(2\pi)^3}}
\newcommand{\dkcp}{\frac{\derivd\k'}{(2\pi)^3}}
\newcommand{\dqc}{\frac{\derivd\p}{(2\pi)^3}}
\newcommand{\dqcp}{\frac{\derivd\p'}{(2\pi)^3}}
\newcommand{\dqcpp}{\frac{\derivd\p''}{(2\pi)^3}}
\newcommand{\be}{\begin{equation}}
\newcommand{\ee}{\end{equation}}
\newcommand{\barr}{\begin{align}}
\newcommand{\earr}{\end{align}}

\newcommand{\cyc}{\ \text{cyc.}}

\newcommand{\Mpc}{\ \text{Mpc}}

\newcommand{\elnn}{\nonumber\\}

\newcommand{\derivd}{\,\mathrm{d}} 

\newcommand{\lin}{{\rm lin}}
\newcommand{\nonlin}{{\rm nonlin}}
\newcommand{\ini}{{\rm in}}
\newcommand{\infl}{{\rm inf}}
\newcommand{\G}{{\rm G}}

\newcommand{\eff}{{\rm eff}}

\newcommand{\E}{{\rm E}}
\newcommand{\La}{{\rm L}}
\newcommand{\LV}{{\rm LV}}
\newcommand{\PS}{{\rm PS}}
\newcommand{\ST}{{\rm ST}}

\newcommand{\R}{{\rm R}}

\newcommand{\vs}{\varphi_{\G,s}}
\newcommand{\vl}{\varphi_{\G,l}}

\title{Non-local bias in the halo bispectrum with primordial non-Gaussianity}

\author[a]{Matteo Tellarini,}
\author[a,b]{Ashley J. Ross,}
\author[c]{Gianmassimo Tasinato}
\author[a]{and David Wands}

\emailAdd{matteo.tellarini@port.ac.uk}
\emailAdd{ross.1333@osu.edu}
\emailAdd{g.tasinato@swansea.ac.uk}
\emailAdd{david.wands@port.ac.uk}

\affiliation[a]{Institute of Cosmology \& Gravitation, University of Portsmouth, Dennis Sciama Building, Portsmouth, PO1 3FX, United Kingdom}
\affiliation[b]{Center for Cosmology \& AstroParticle Physics, The Ohio State University, Columbus, OH 43210, USA}
\affiliation[c]{Department of Physics, Swansea University, Swansea, SA2 8PP, UK}

\abstract{
Primordial non-Gaussianity can lead to a scale-dependent bias in the density of collapsed halos relative to the underlying matter density. 
The galaxy power spectrum already provides constraints on local-type primordial non-Gaussianity complementary those from the cosmic microwave background (CMB), while the bispectrum contains additional shape information and has the potential to outperform CMB constraints in future. 
We develop the bias model for the halo density contrast in the presence of local-type primordial non-Gaussianity, deriving a bivariate expansion up to second order in terms of the local linear matter density contrast and the local gravitational potential in Lagrangian coordinates.
Nonlinear evolution of the matter density introduces a non-local tidal term in the halo model. Furthermore,  the presence of local-type non-Gaussianity in the Lagrangian frame leads to a novel non-local convective term in the Eulerian frame,  that is proportional to the displacement field when going beyond the spherical collapse approximation.
We use an extended Press-Schechter approach to evaluate the halo mass function and thus the halo bispectrum. We show that including these non-local terms in the halo bispectra can lead to corrections of up to $25\%$ for some configurations, on large scales or at high redshift.
}

\keywords{Primordial non-Gaussianity, Large Scale Structure}

\arxivnumber{1504.00324}

\begin{document}
\maketitle
\section{Introduction}\label{sec:intro}
 
  \subsection{General motivations}
  Inflation is a powerful mechanism to solve the puzzle of initial conditions for the Hot Big Bang cosmology and explain the origin of structure in our Universe. However, the exact mechanism that drove inflation is still a matter of debate. Different models make different predictions for the statistics of the primordial curvature perturbations $\zeta_\infl$, which can be quantified by computing the $n$-point correlation functions $\langle \zeta_\infl (\k_1) \dots \zeta_\infl (\k_n) \rangle$. 
All models predict some departure from Gaussianity, giving non-vanishing correlation functions for $n>2$. Since the amplitude, shape- and scale-dependence are model dependent, the intrinsic non-Gaussianity of primordial curvature perturbations is a powerful tool to discriminate among different models \cite{Bartolo:2004if}.
For example, local-type non Gaussianity is a distinctive signature of multi-field models \cite{Wands-2010review}, while equilateral and orthogonal non-Gaussianity probe non-slow roll dynamics through the self-interactions of the fields \cite{Koyama:2010xj}. 
Thus observational features of the primordial curvature perturbation offer a window onto the nature of inflation \cite{Alvarez:2014vva}. 

At the time of this publication, the most accurate constraints on primordial non-Gaussianity (PNG) come from the cosmic microwave background (CMB). WMAP$9$ set the constraint on local $\fnl$ to be $-3<\fnl<77$ at $95\%$ CL \cite{WMAP9}, while the most recent data from the Planck satellite experiment requires $-9.2<\fnl<10.8$ \cite{Ade:2015ava}. 
  This is compatible with the predictions of simple slowly-rolling single-field models which predict $|\fnl|<1$.     
  The CMB is a $2$D map at redshift $z \approx 1100$ whose temperature fluctuations have been measured with high accuracy. The error bars could be further decreased by including the higher-resolution polarization data \cite{Andre:2013nfa}, but we are close to having fully exploited the constraining power of the CMB, being limited on small scales by the Silk damping and on large scales by the cosmic variance. The question is then which other observables may be suitable to obtain bounds on $\Delta \fnl \approx 1$.

  The large-scale structure (LSS) of the Universe  provides a $3$D map of the Universe, with accessible modes potentially going from the horizon scale $\sim 10^4 \Mpc$ down to the non-linear scale $\sim 10 \Mpc$. 
  The pioneering paper by Dalal et al. \cite{Dalal:2007cu} (see also \cite{Matarrese:2008nc})  showed that the LSS power spectrum also offers a strong constraint on primordial non-Gaussianity.
  The idea comes from the mode coupling in local-type models; the short wavelength modes which drive the collapse of matter into halos are modulated by the long wavelength modes, effectively changing the short-mode variance from patch to patch of the sky. This introduces a scale-dependent correction in the bias model, which describes how the number of halos relates to the matter density.
  Since the correction is proportional to $\fnl / k^2$, the clustering of structures on large scales has the potential to discriminate among different models of inflation.

It is important to emphasise that LSS constraints are independent of those from the CMB and apply on different scales, hence constraining also scale-dependent non-Gaussianity.
This new perspective on the subject  has been studied in depth in a number of works \cite{Matarrese:2008nc,Slosar:2008hx,Afshordi:2008ru,Desjacques:2008vf,Desjacques1,Desjacques2,Gong}. However, constraining $\fnl$ from real data requires detailed understanding of the systematic errors; indeed, some of them mimic the PNG excess of power on large scales (see for instance \cite{Huterer2013,Pullen2013,Agarwal:2013ajb}). Different techniques have been proposed to handle these errors \cite{Pullen2013,Leistedt:2014zqa,Ross:2013,Karagiannis:2013xea}, as well as methods to reduce the effect of cosmic variance \cite{Seljak:2008xr,GilMarin10,Hamaus11,Biagetti13,Ferramacho:2014pua,Yamauchi:2014ioa} and the statistical uncertainties  \cite{Seljak09,Hamaus10}. Constraints comparable with the WMAP bounds have already been achieved \cite{Slosar:2008hx,Xia1:2010,Xia2:2010,Ross:2013,Karagiannis:2013xea,Giannantonio:2013uqa,Ho:2013lda,Giannantonio:2013kqa,Leistedt:2014zqa} and even more stringent results will come thanks to the next generation of experiments, like Euclid and SKA \cite{Giannantonio:2011,Raccanelli:2014kga,Camera:2014bwa,Yamauchi:2014ioa}, despite the fact that no survey has been optimized for PGN so far \cite{dePutter:2014lna}. The novel technique introduced in \cite{Leistedt:2014wia} may optimistically provide $\Delta \fnl \sim 1$.

  However the $2$-point function is not the natural statistic in which to look for PNG, as the full shape information is available only in higher-order statistics, and moreover local-type models including higher-order non-Gaussian parameters like $\gnl$ are approximately degenerate with $\fnl$ in the power spectrum \cite{Roth:2012}.  
  These issues naturally drive us towards studying higher-order $n$-points correlation functions, in particular the $3$-point function and its Fourier transform, the \emph{bispectrum}\footnote{In \cite{Mao:2014caa} the possibility to use the higher-order moments of LSS to constraint PNG has been explored. Although these avoid the complexities of the full $n$-point statistics, they may not be able to detect small PNG.}.  Many more $3$D triangle configurations compared to $2$D CMB ones are available in $k$-space, suggesting potentially a stronger constraining power with respect to the power spectrum. Indeed, $\Delta \fnl \approx 1$ is expected from the bispectrum \cite{Scoccimarro:2003wn,Sefusatti:2007ih,Baldauf:2010vn}, although this forecast is based on idealized surveys and ignores  redshift space distortions (RSD). 
  In addition, the degeneracy between $\fnl$ and $\gnl$ is broken by combining both the power spectrum and bispectrum constraints \cite{Jeong:2009,Tasinato:2013vna}.

  There are a number of complications, which explains why few measurements of the $3$-point function or bispectrum from real data have been obtained so far \cite{Scoccimarro:2000sp,Verde:2001sf,Scoccimarro:2003wn,Jing:2003nb,Gaztanaga:2005an,FMarin:2011,McBride:2011,Marin:2013bbb,Gil-Marin:2014sta,Gil-Marin:2014baa}. 
  For example, redshift-space distortions and the complicated mask geometry intrinsic to any survey are hard to model. Further, non-linearities in halo populations and the non-linear nature of Einstein’s field equations produce non-Gaussian features in the evolved matter field that may be hard to disentangle from the PNG signal.

  \subsection{What we do in this work}
  The ability to go beyond current constraints on PNG, crucially depends on our understanding of all possible effects that can influence LSS measurements and how these depend on the primordial fluctuations.
  In this paper we make improvements to the bias model which relates the halo density to the underlying matter density, particularly focussing on an accurate description of the second-order, non-local and non-Gaussian effects.

  A common procedure to describe how dark matter halos trace the matter density is to assume a local relation between the final number density of objects with mass $\MA$ at redshift $z$ and the evolved density field, known as \emph{local Eulerian biasing} \cite{Fry:1992vr},
  \begin{equation}\label{eq:local_eulerian}
    \delta_h^\E(\x,z,\MA) = \sum_{j=1}^{\infty} \, \frac{b^\E_j(z)}{j!} \left( \delta^\E_\MA (\x,z) \right)^j \,,
  \end{equation}
  where $b_j$ are the bias coefficients and $\delta^\E_\MA$ is the fully non-linear density contrast in Eulerian space, smoothed on the mass scale $\MA$. 
  In this picture, the halos are drawn on top of the density field without any memory of their past history, with possible application also to the case of PNG \cite{Scoccimarro:2003wn,Sefusatti:2007ih,Jeong:2009}. However, many recent results show that this biasing model is not sufficiently accurate  when compared to simulations \cite{Baldauf:2012hs,Saito:2014qha,Chan:2012,Roth:2011,Pollack:2013alj}. In particular,  other physically motivated contributions consistent with the symmetries of the dark matter equations of motion \cite{2009JCAP...08..020M}, like a tidal term, should be included. These new terms break the assumption of locality implicit in \cref{eq:local_eulerian}.
  Either a local or non-local Eulerian model offer only a parametrization of the bias, and the bias coefficients need to be fitted against data or N-body simulations.

  A promising  alternative is to assume a physically reasonable local relation between the {\it initial} number density of objects and the {\it initial} density field \cite{Mo:1995cs},
  \begin{equation}\label{eq:local_lagrangian}
    \delta_h^\La(\q,z,\MA) = \sum_{j=1}^{\infty} \, \frac{b^\La_j(z)}{j!} \left( \delta^\La_{\lin,\MA} (\q,z) \right)^j \,,
  \end{equation}
  where $\delta^\La_{\lin,\MA}$ is the linearly evolving density contrast smoothed on scale $\MA$. Note that the expansion is performed in Lagrangian space, with the initial spatial coordinate $\q$ being related to the evolved Eulerian coordinate, $\x$, through  the relation 
  \begin{equation}\label{eq:lag_to_eu}
    \x(\q,\tau)=\q+\ps(\q,\tau) \,,
  \end{equation}
  where $\ps$, the displacement field,   is the key dynamical quantity in the Lagrangian picture.
  \Cref{eq:local_lagrangian} is known  as a \emph{local Lagrangian biasing} scheme and in effect assumes that the formation sites of halos can be identified from the initial density field~\footnote{Generalizations of this picture exists as well, see for instance \cite{Sheth:2013,Matsubara:2011PhRvD}.}. The dynamics of halos is then captured in the transformation to the Eulerian space, by applying \cref{eq:lag_to_eu}.
  This local Lagrangian biasing scheme comes with a prescription to calculate the bias, if we know  the halo mass function, and it also {\it automatically}  leads to the presence of a tidal term when written in terms of the non-linearly evolved density field \cite{Catelan:2000vn,Baldauf:2012hs}.
  We will follow this approach, carefully evaluating the effects of the displacement field on the physical quantities we will  investigate.

  In presence of PNG of local type, the usual local Lagrangian bias model needs to be extended to account for the correlations in the initial density field. In the following we start re-deriving the bivariate model of \cite{Giannantonio:2009ak},  which remarkably shows good fits against simulations even when applied to the bispectrum \cite{Baldauf:2010vn,Sefusatti:2011}. 
  We then show that a novel non-local term arises, contributing to the number density of halos, which was previously neglected under the assumption of spherical collapse.  We will  quantitatively investigate   how much the tidal term,  and our  new non-local contribution, affect the tree-level halo bispectrum in comparison with the reference model of \cite{Baldauf:2010vn} (hereafter BSS)\footnote{Alternative frameworks to include PNG of general type in a Lagrangian description exist as well. See for instance \cite{Scoccimarro:2012} and \cite{Yokoyama:2013mta}, which is based on the integrated perturbation theory developed in \cite{Matsubara:2011PhRvD}}. 

\bigskip

This paper is organized as follows. In \cref{sec:basics}, we briefly introduce the main concepts and notation. In particular, 
in \cref{subsec:densityfield} we review the evolution of the matter density field up to second order and indentify a new convective term in presence of PNG, while in \cref{subsec:pbs} we describe the peak-background split. This  constitutes the basis of our local Lagrangian biasing scheme, which will be presented in \cref{sec:lagrangian}. As we work in a Lagrangian approach, our result then needs to be transformed into the Eulerian frame in \cref{sec:eulerian}. In \cref{sec:bispectrum} expressions for the matter and halo bispectra are given, and a numerical analysis of the halo bispectrum is presented in \cref{sec:analysis}, where we quantify the effect of new non-local terms in LSS bispectra. We present our final conclusions together with a discussion of potential future developments of our work in \cref{sec:discussion}.  

\section{Basic framework}\label{sec:basics}

  \subsection{Matter density field} \label{subsec:densityfield}

In this subsection we define the quantities and introduce the notation that we will use in what follows. 
The linearly growing mode of the density field has a simple separable form in Lagrangian space
\begin{equation} 
\delta_\lin (\q,z) = C(\q) D(z) \,,
\end{equation}
where $D(z)$ is the linear growth factor with  $D(z) \propto (1+z)^{-1}$ in the matter-dominated era, normalized such that $D(0)=1$.

\begin{figure}
\centering
\subfloat
{\includegraphics[width=.45\textwidth]{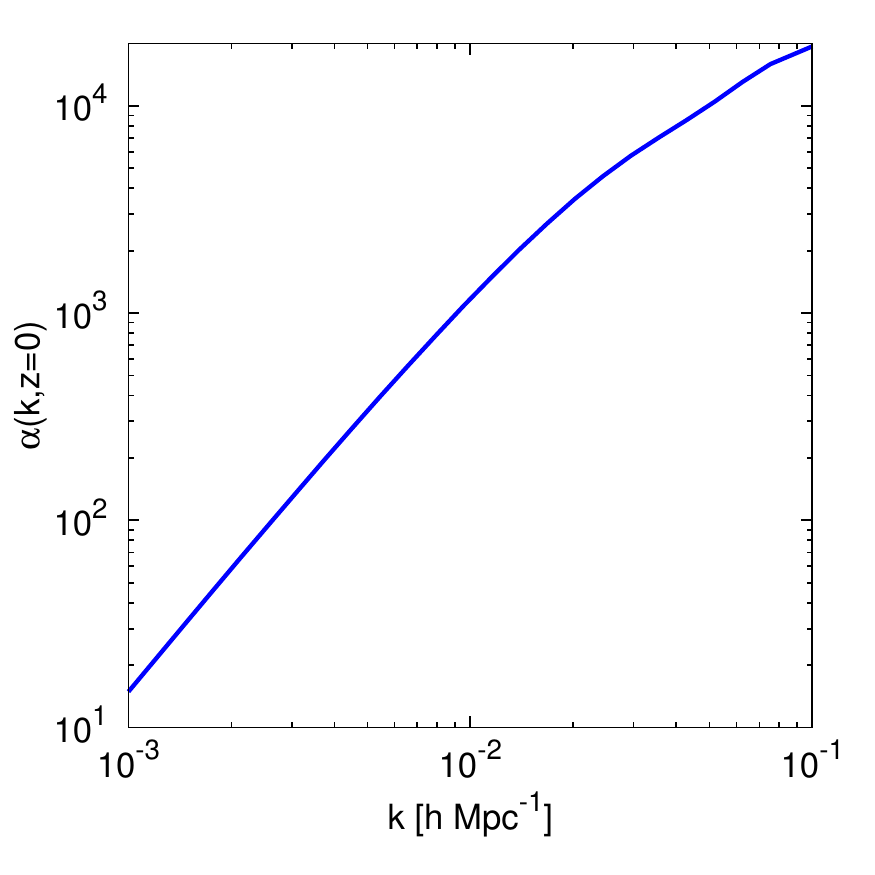}} \quad
\subfloat
{\includegraphics[width=.45\textwidth]{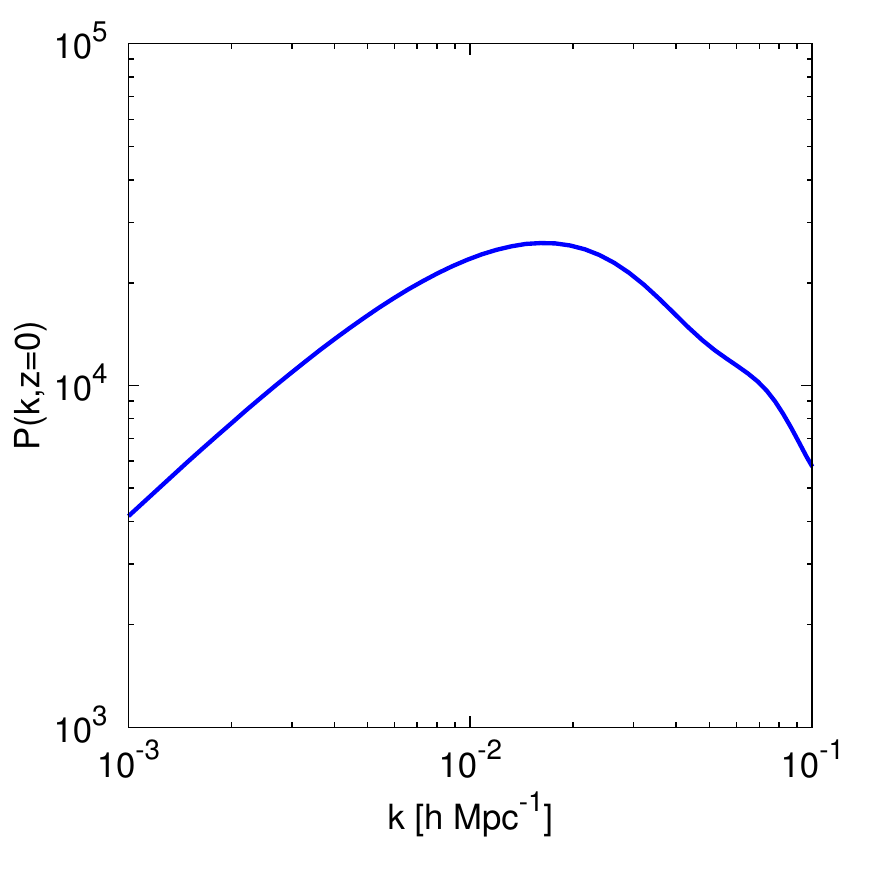}}
\caption{The left panel shows $\alpha$ as a function of wavenumber $k$ at redshift $z=0$. The right panel shows the matter power spectrum $P(k)$ at redshift $z=0$. They are both obtained using CAMB \cite{camb} and assuming the Planck data set \cite{Ade:2013ydc}.}
\label{fig:alpha&p}
\end{figure}

The density contrast at the start of the matter-dominated era is determined by the primordial metric perturbation from inflation
 that we denote with  $\zeta_\infl$.  
In Fourier space we have
\begin{equation} 
 \label{eq:zetainfl}
 \delta_\lin (\k,z) = \frac{2 k^2 c^2 T(k) D(z)}{5 \Omega_m H_0^2} \, \zeta_\infl(\k) \,,
\end{equation}
where the transfer function $T(k)$ accounts for the damping of sub-Hubble-scale modes in the radiation era and $T(k) \rightarrow 1$ as $k\to0$.  Conventionally this is written in terms of a primordial Newtonian potential
\begin{equation}
 \label{eq:alpha}
 \delta_\lin (\k,z) = \alpha(k,z) \Phi_\ini (\k) \,,
\end{equation}
where, from \cref{eq:alpha}, we identify 
\begin{equation}
\label{eq:Phi}
\Phi_\ini = \frac35 \zeta_\infl \, , \quad
\alpha(k,z) \equiv  \frac{2 k^2 c^2 T(k) D(z)}{3 \Omega_m H_0^2} \,.
\end{equation}
The function $\alpha$ is plotted in \cref{fig:alpha&p}. 

Throughout this paper we will consider a primordial Newtonian potential with local-type non-Gaussianity, that is a local function of a Gaussian random field \cite{Gangui:1993tt,Verde:1999ij,Komatsu:2001rj}
\begin{equation}\label{eq:png}
 \Phi_\ini = \vp_\G (\q) + \fnl \left( \vp^2_\G (\q) - \langle \vp^2_\G \rangle \right) \,,
\end{equation}
where $\vp_G(\q)$ is a Gaussian field, seeded by free field fluctuations during inflation, and $\fnl$ is a dimensionless parameter quantifying the magnitude of non-Gaussian corrections either due to non-linear evolution of the primordial metric perturbation \cite{Wands-2010review}, or
 to the  non-linearity inherent in the general relativistic constraint equations \cite{Bartolo2010,Bruni:2013qta,Bruni:2014xma}. Note that the non-Gaussian correction is a local function of the Gaussian field at a given initial (Lagrangian) position, $\q$.

In the presence of primordial non-Gaussianity, the linearly growing mode (\ref{eq:alpha}) contains first- and second-order terms with respect to  $\vp_G(\q)$.  It will be convenient, for our discussion,  to define the first-order (Gaussian) density contrast $\delta_{\G}(\k,z)$, where from \cref{eq:alpha} and \cref{eq:png} we identify
\begin{equation}
 \label{eq:deltaG}
 \delta_{\G} (\k,z) = \alpha (k,z) \vp_\G (\k) \,.
\end{equation}

The initial linearly growing mode of \cref{eq:alpha} drives the subsequent non-linear evolution of the density field \cite{Bernardeau:2001qr} and we may write the non-linearly evolved field in Lagrangian coordinates as
	\begin{align}\label{eq:delta_q_lin_nonlin}
	  \delta & = \delta_\lin (\q,z) + \delta_\nonlin^\La (\q,z) \,.  
	\end{align}
where at second order we have
 	\begin{align}
         \delta^\La_\nonlin (\q,z) & \simeq \frac{17}{21} (\delta_\lin (\q,z))^2 + \frac{2}{7} s^2(\q,z)  \label{eq:secL} \,
	\end{align}
with $s^2=s_{ij}s^{ij}$ and $s_{ij}$, the trace-free \emph{tidal tensor}, is defined as
	\begin{equation}
	  s_{ij} \equiv \left( \nabla_i\nabla_j -  \frac{1}{3} \delta^K_{ij} \right) \nabla^{-2}\delta \,,
	\end{equation} 
	where $\delta^K_{ij}$ is the Kronecker delta function.

Since the Lagrangian and Eulerian coordinates (\ref{eq:lag_to_eu}) agree at leading order, the first-order density perturbation \cref{eq:deltaG} has the same form in either frame. However at second order the density perturbations differ because of the convective term proportional to the displacement, $\ps$ \cite{Bertacca:2015mca}\footnote{See also \cite{Schmittfull:2014tca} where this term is referred to as a \emph{shift} term.}. For a general function $f(\x,z)$ we have, from \cref{eq:lag_to_eu},
\begin{equation}
 \label{convective}
 f(\x(\q,z),z) \simeq f(\q,z) + \ps \cdot \na f(\q,z) \,.
\end{equation}
Hence we can write the non-linearly evolved density field \cref{eq:delta_q_lin_nonlin} in Eulerian coordinates as
	\begin{align}
	  \delta & = \delta_\lin (\x,z) + \delta_\nonlin^\E (\x,z) \label{eq:delta_x_lin_nonlin} \,,
	\end{align}
where at second order, using \cref{eq:secL,convective}, we obtain  \cite{peebles1980large}:
 	\begin{align} 
	\label{eq:secE}
	\delta^\E_\nonlin (\x,z) & \simeq \frac{17}{21} (\delta_\lin (\x,z))^2 + \frac{2}{7} s^2(\x,z) - \ps(\x,z) \cdot \na\delta(\x,z) \,.
	\end{align}
Similarly, although the Newtonian potential has the same form to first order, we must include a convective term when writing the primordial potential (\ref{eq:png}) up to second order in Euclidean coordinates
\begin{equation}
\label{eq:varphitransf}
 \vp_\G (\x) \simeq \vp_\G (\q) + \ps (\x,z) \cdot \nabla \vp_\G (\q) \,;
\end{equation}
and thus
\begin{equation}
\Phi_\ini \simeq \vp_G(\x) - \ps(\x,z) \cdot \na\vp_G(\x) + \fnl \left( \vp^2_\G (\x) - \langle \vp^2_\G \rangle \right) \,.
\end{equation}

Hence we learn that, even though the primordial potential is by definition a local function of an initial Gaussian random field $\vp_G(\q)$ at each initial spatial coordinate $\q$ and at a fixed initial time, the primordial potential at a fixed Eulerian position, $\x$  becomes time-dependent at second-order,  due to the time-dependent relation between Eulerian and Lagrangian coordinates (\ref{eq:lag_to_eu}). 
In \cref{sec:displaced} we show that even a Gaussian initial potential in the Lagrangian frame becomes a non-Gaussian field at second-order in the Eulerian frame due to the first-order displacement.
We shall see that this gives rise to additional non-local terms in the distribution of collapsed halos in Eulerian space in the presence of primordial non-Gaussianity, assuming a bivariate local bias model (dependent on the linear density and the primordial potential) in Lagrangian space.


\subsection{Peak-Background split}\label{subsec:pbs}

Our local Lagrangian bias model relies on a simple though powerful argument, known as the peak-background split. In this approach,  the first-order potential $\vp_\G$ is considered as a superposition of long and short modes
\begin{equation}
 \label{eq:pbs}
\vp_\G(\q) = \vl(\q) + \vs(\q) \,, 
\end{equation}
which are statistically independent for a Gaussian random field. The local background is composed of long-wavelength modes $\vl$
and acts as an approximately homogeneous background cosmology on comoving scale $l$. On top of these the smaller scale peaks $\vs$
lead to the collapse of dark matter into halos, on a scale $\R\ll l$, 
when exceeding a suitable threshold value, $\delta_c$. This is usually assumed to be the linearly growing density mode for a spherically collapsed object ($\delta_c=1.686$).

With Gaussian initial conditions, this implies that the threshold for collapse is effectively different from place to place,
\begin{equation}
  \delta_c \longrightarrow \delta_c - \delta_{\G,l}(\q) \, ,
\end{equation}
enhancing the formation of structures on top of long-mode overdense regions; this conclusion leads naturally to the idea that dark matter halos are biased tracers of the underlying matter distribution on a given scale $l$ \cite{Kaiser:1984sw}. 

Substituting the peak-background split (\cref{eq:pbs}) into the non-Gaussian primordial potential of \cref{eq:png}, we obtain
\begin{equation}\label{eq:pbspng}
    \Phi_\ini (\q) = \vl + \fnl \left( \vl^2 - \langle \vl^2 \rangle \right) + \left( 1 + 2 \fnl \vl \right) \vs + \fnl \left( \vs^2 - [\vs^2]_V \right) \,,
\end{equation}
where the square brackets $[\vs^2]_V$ account for a local expectation evaluated over the volume $V \sim l^3$ centred around $\q$, 
\begin{equation}
  [\vs^2]_V (\q) \equiv \int_{k \geq l^{-1}} \dk \int_{k' \geq l^{-1}} \dkcp e^{-i(\k+\k')\cdot\q} \langle \vs(\k) \vs(\k') \rangle  \, .
\end{equation}

By Fourier transforming \cref{eq:pbspng} and defining the Gaussian long and short density mode respectively as $\delta_{\lin,\G,l}=\alpha \, \vl$ and $\delta_{\lin,\G,s}=\alpha \, \vs$, we can identify a ``background'' density perturbation
\begin{equation}
  \delta_{\lin,l} (\k) = \delta_{\G,l} + \fnl \alpha \left( \vl^2 - \langle \vl^2 \rangle \right)\,,
\end{equation}
which changes the effective collapse threshold for the small scale peaks
\begin{equation}
\delta_c \, \longrightarrow \, \delta_c - \delta_{\lin,l} \, .
\end{equation}
At the same time, local-type non-Gaussianity introduces a correlation between long and short wavelength modes in the density field. 
This leads to the important conclusion that the small scale power is affected by the long-modes of the primordial potential,
\begin{equation}
 \label{eq:sigmal}
  \sigma_l = (1 + 2 \fnl \vl) \sigma_G \,.
\end{equation}

The above discussion suggests that the number density of objects with mass $\MA$ at redshift $z$ in a volume $V\sim l^3$, i.e. the local \emph{mass function} (see \cref{app:massfunction} for a brief introduction), depends not only on the mass and the redshift but also on the local density field and its moments \cite{Baumann:2013,Tasinato:2013vna}
\begin{equation}\label{eq:pngnh}
  n_h = n_h(\MA,z, [\delta^n_\lin]_V) \,.
\end{equation}
In the non-Gaussian cosmology of \cref{eq:png}, \cref{eq:pngnh} for a given halo mass $\MA$ and redshift $z$ simply reduces to $n_h = n_h (\delta_{\lin,l},\sigma_l,\fnl)$.
The dimensionless variable $\nu$ in the mass function (\cref{eq:nu}) thus needs to be replaced with the local effective value
\begin{equation}
\label{eq:nul}
  \nu = \frac{\delta_c}{\sigma_G} \, \longrightarrow \, \nu(\q) = \frac{\delta_c - \delta_{\lin,l}(\q)}{\sigma_l (\q)} \, .
\end{equation}
This is the main result of this section and provides the basis for our local bias model in Lagrangian space.


\section{Local Lagrangian bias model: halo overdensity}
\label{sec:lagrangian}



Our bias model provides an expression for the local halo overdensity in Lagrangian coordinates
\begin{equation}
 \label{eq:defdh}
 \delta_h^\La (\q) = \frac{n_h(\q) - \langle n_h \rangle}{\langle n_h \rangle} \, .
\end{equation}
Using the preceding arguments we expect the local halo overdensity to be a function of the large-scale linearly growing mode of the density field, $\delta_{\lin,l}$ (the local ``background'' density), and the small-scale variance of the linearly growing mode, $\sigma_l$ (the ``peaks''), averaged over the same large scale ($l$ in the preceding section).
We thus Taylor expand \cref{eq:pngnh} up to second order\footnote{Note that to compute the tree-level bispectrum we need quantities up to $\mathcal{O}(\delta^2)$.} in terms of $\delta_{\lin,l}$ and $\sigma_l$ 
\begin{equation}
\begin{split}
 \delta_h^{\La}(\q)=\,& \beta_{10} \delta_{\lin,l} + \beta_{01} \left(\frac{\sigma_l}{\sigma}-1\right) +\\&+\frac{1}{2}\left[\beta_{20} (\delta_{\lin,l})^2 + \beta_{02} \left(\frac{\sigma_l}{\sigma}-1\right)^2 + 2 \beta_{11} \delta_{\lin,l} \left(\frac{\sigma_l}{\sigma}-1\right) \right] \,,
\end{split}
\end{equation}
where we define the bias coefficients
\begin{equation}
 \label{eq:betaij}
 \beta_{ij} \equiv \left[\frac{(\sigma_l)^j}{n_h}\frac{\partial^{i+j}n_h}{\partial^i\delta_{\lin,l} \partial^j \sigma_l} \right] \biggr\vert_{\delta_{\lin,l}=0, \sigma_l=\sigma} \, .
\end{equation}
Hereafter we drop the subscript $l$,  and simply use $\delta_\lin$ and $\vp_\G$ to indicate the long-wavelength modes of the linearly growing density contrast and the Gaussian primordial potential respectively.

The small-scale variance of \cref{eq:sigmal} depends on the local primordial potential $\vp_{\G,l}$ and hence we can write the Taylor expansion in the bivariate form \cite{Giannantonio:2009ak,Baldauf:2010vn}
\begin{equation}\label{eq:bivariate}
 \delta^\La_h(\q)=\, b^\La_{10} \delta_{\lin} + b^\La_{01} \vp_\G + b^\La_{20} (\delta_{\lin})^2 + b^\La_{11} \delta_{\lin} \vp_\G + b^\La_{02} \vp_\G^2 \, ,
\end{equation}
where we identify the Lagrangian bias coefficients
\begin{equation}\label{eq:newlagrangianbias}
  \begin{split}
    b_{10}^\La & = \beta_{10} \,, \\
    b_{01}^\La & = 2 \fnl \beta_{01}  \,,\\
    b_{20}^\La & = \frac{\beta_{20}}{2}  \,,\\
    b_{11}^\La & = 2 \fnl \beta_{11}  \,,\\
    b_{02}^\La & = 2 \fnl^2 \beta_{02} \,,
  \end{split}
\end{equation}

Following BSS, we assume the Sheth-Tormen mass function corrected for non-Gaussian initial conditions (see \cref{app:massfunction} for further details). This allows to provide explicit formulas for the Lagrangian bias coefficients; only two of them are independent:
\begin{align}
      b_{10}^\La & = \frac{\gamma \nu^2 -1}{\delta_c} + \frac{2p}{1+(\gamma \nu^2)^p}\frac{1}{\delta_c} - \kt \frac{\nu^3 - \nu}{2\delta_c} + \dkt \frac{\nu + \nu^{-1}}{6 \delta_c} \label{eq:b10}\\
      b_{20}^\La & = \gamma \nu^2 \frac{\gamma \nu^2 - 3}{2 \delta_c^2} + \frac{p}{1+(\gamma \nu^2)^p} \frac{2\gamma\nu^2 + 2p -1}{\delta_c^2} - \frac{\kt}{2} \left[ \frac{\gamma \nu^5 -(\gamma+2)\nu^3 + \nu}{\delta_c^2} + \frac{2p}{1+(\gamma \nu^2)^p} \frac{\nu^3 - \nu}{\delta_c^2}\right] + \nonumber \\ & \quad + \frac{1}{2} \dkt \left[ \frac{\gamma\nu^3 + (\gamma -1) \nu}{3\delta_c^2} + \frac{2p}{1+(\gamma \nu^2)^p} \frac{\nu - \nu^{-1}}{3\delta_c^2} \right] \,, \label{eq:b20} 
\end{align}
where $\gamma$ and $p$ were introduced in \cref{eq:st}.
%
%
Thus the remaining coefficients are obtained from the following combinations
\begin{equation}
  \begin{split}
    b_{01}^\La & = 2 \fnl \delta_c b_{10}^\La  \,,\\
    b_{11}^\La & = 2 \fnl (\delta_c b_{20}^\La - b_{10}^\La)  \,,\\
    b_{02}^\La & = 4 \fnl^2 \delta_c (\delta_c b_{20}^\La - 2b_{10}^\La)  \,.  
  \end{split}
\end{equation}
In \cref{eq:b10} we can identify the first two terms as the usual Gaussian bias \cite{Scoccimarro:2000gm} plus a scale-independent correction introduced by PNG \cite{Slosar:2008hx,Afshordi:2008ru}, which has been showed to improve the comparison between theory and simulations \cite{Desjacques:2008vf}. The same structure is found for \cref{eq:b20}.

\section{Non-local Eulerian bias}\label{sec:eulerian}

Galaxy surveys map the distribution of galaxies, which we assume to be located in collapsed halos, displaced with respect to their Lagrangian positions according to \cref{eq:lag_to_eu}. \Cref{eq:bivariate} describes the excess of halos in Lagrangian space but this needs to be transformed to the Eulerian frame to account for their dynamics.

Unlike the matter density contrast, the halo density contrast is not a $3$-scalar since it is conventionally defined as a coordinate density \cite{Bertacca:2015mca}. Hence the number of halos in a given volume element is given by
\begin{equation}
N_h  = n_h^\La (\q,z) d^3\q = n_h^\E (\x,z) d^3\x \,.
\end{equation} 
Thus we have
\begin{equation}
1+\delta_h^\E (\x,z) = \left[ 1+ \delta_h^\La (\q,z) \right] \left| \frac{d^3\q}{d^3\x} \right|
\end{equation} 
and using the coordinate Jacobian [see \cref{app:transf}, \cref{eq:J}] we obtain the transformation rule \cite{Catelan:1997qw,Matsubara:2011PhRvD}
\begin{equation}
\label{eq:nhE}
1+\delta_h^\E (\x,z) = [1+\delta(\x,z)] \left[ 1+ \delta_h^\La (\q,z) \right] \,,
\end{equation} 

The Lagrangian space halo density contrast, $\delta_h^\La (\q,z)$, is given by \cref{eq:bivariate} in terms of the linearly growing density contrast $\delta_\lin(\q,z)$ and the Gaussian potential $\vp_G(\q)$ in Lagrangian coordinates. However,  we have seen in \cref{eq:delta_q_lin_nonlin,eq:secL} how $\delta_{\lin}(\q,z)$  can be expressed up to second-order in terms of the non-linear matter density, $\delta$, and the tidal tensor, $s^2$,
\begin{equation}
\label{eq:linLE}
\delta_\lin (\q,z)  \simeq  \delta (\x,z) - \frac{17}{21}(\delta (\x,z))^2 - \frac27 s^2 \,.
\end{equation}
 In \cref{eq:varphitransf} we showed how $\vp_G(\q)$ can be expressed up to second-order in terms of the displacement $\ps$ and the Gaussian potential in Eulerian coordinates $\vp_G(\x)$:
\begin{equation}
 \vp_\G (\q) \simeq \vp_\G (\x) - \ps (\x,z) \cdot \nabla \vp_\G (\x) \,. 
\end{equation}

\begin{figure}
\centering
\includegraphics[width=1.0\textwidth]{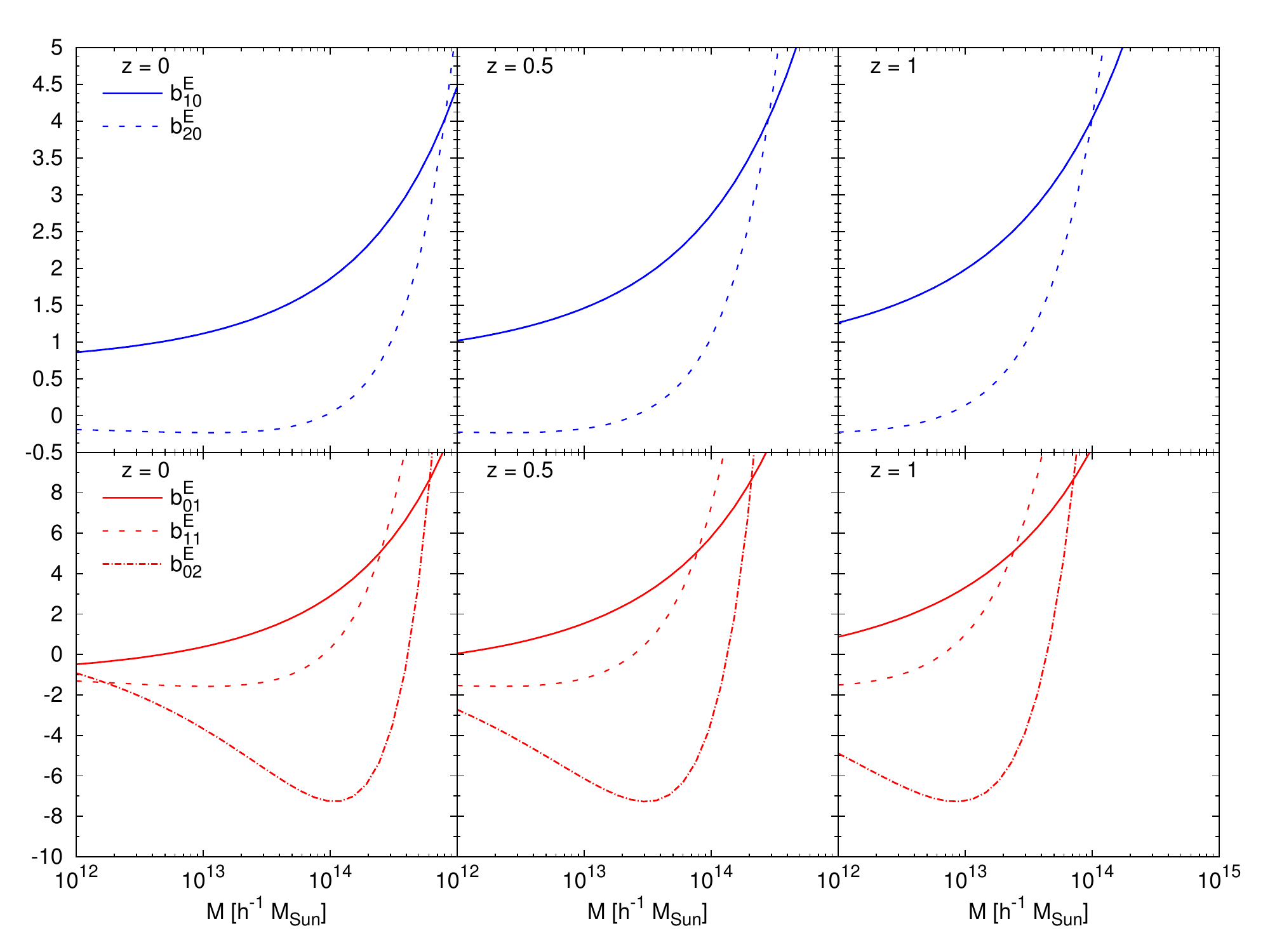}
\caption{The Eulerian bias coefficients as a function of mass and redshift, assuming $\fnl=1$.}
\label{fig:b10b20}
\end{figure}

We thus obtain our expression for the Eulerian halo overdensity up to second order in terms of the density contrast and the Gaussian potential in Eulerian coordinates
\begin{equation}
\begin{split}\label{eq:deltahx}
  \delta^\E_h(\x)=\,& b_{10}^\E \delta + b_{01}^\E \varphi + b_{20}^\E \delta^2 + b_{11}^\E \vp_\G \delta + b_{02}^\E \vp_\G^2 - \frac{2}{7} b_{10}^\La s^2 - b_{01}^\La \ps(\x,z) \cdot \na \vp_\G \,.
\end{split}
\end{equation}
where we define the standard Eulerian bias coefficients (see \cref{fig:b10b20}),
\begin{equation}\label{eq:eul_bias}
\begin{split}
 b_{10}^\E & = 1+b_{10}^\La \\
 b_{01}^\E & = b_{01}^\La \\
 b_{20}^\E & = \frac{8}{21} b_{10}^\La + b_{20}^\La \\
 b_{11}^\E & = b_{01}^\La + b_{11}^\La \\
 b_{02}^\E & = b_{02}^\La \, ,
\end{split}
\end{equation}
{\Cref{eq:deltahx} generalises the result that is usually obtained under the spherical collapse approximation (see \cref{app:local_eul_b})}. Indeed, the last two terms of \cref{eq:deltahx} are the non-local, non-linear terms of our bias model. While $s^2$ is an already known tidal term, we have derived here for the first time the convective contribution $\ps(\x,z) \cdot \na \vp_\G$. We decide to keep the corresponding bias coefficients written as $b_{10}^\La$ and $b_{01}^\La$, instead of replacing them with $b_{10}^\E -1 $ and $b_{01}^\E$ respectively. In this way we will be able to recognise more easily the differences they introduce with respect to the reference model of BSS, especially in the discussion in \cref{sec:analysis}.

Finally, if we perform a Fourier transform with respect to Euclidean coordinates we obtain
\begin{equation}
\label{eq:deltahk}
\begin{split}
\delta_h^\E(\k)=\,& b_{10}^\E \delta + b_{01}^\E \vp_\G + b_{20}^\E \delta \ast \delta + b_{11}^\E \delta \ast \vp_\G + b_{02}^\E \vp_\G \ast \vp_\G - \frac{2}{7} b_{10}^\La s^2 - b_{01}^\La n^2 \,,
\end{split}
\end{equation}
where we define
\begin{align}
\label{eq:deltaEk}
\delta^\E (\k)  = & \, \delta_\G(\k) + \fnl \alpha(k) \int \frac{d\q}{(2 \pi)^3} \frac{\delta_\G(\q)\delta_\G(\k-\q)}{\alpha(q) \alpha(\vert \k-\q \vert)} 
+ \int\,\frac{d\q}{(2 \pi)^3} \mathcal{F}_2(\q,\k-\q)\delta_\G(\q)\delta_\G(\k-\q) \\ 
s^2 (\k) = & \, \int \frac{d\q}{(2 \pi)^3} \mathcal{S}_2(\q,\k-\q)\delta_\G(\q)\delta_\G(\k-\q)\\
n^2 (\k) = & \, 2 \int \frac{d\q}{(2 \pi)^3} \mathcal{N}_2(\q,\k-\q)\frac{\delta_\G(\q)\delta_\G(\k-\q)}{\alpha(\vert \k-\q \vert)}
\end{align}
and the kernels are given by
\begin{align}
 \mathcal{F}_2(\k_1,\k_2) & = \, \frac{5}{7} + \frac{1}{2} \frac{\k_1 \cdot \k_2}{k_1 k_2} \left(\frac{k_1}{k_2} + \frac{k_2}{k_1} \right) + \frac{2}{7} \frac{\left( \k_1 \cdot \k_2 \right)^2}{k_1^2 k_2^2} \\
 \mathcal{S}_2(\k_1,\k_2) & = \, \frac{\left( \k_1 \cdot \k_2 \right)^2}{k_1^2 k_2^2} - \frac{1}{3} \\
 \mathcal{N}_2(\k_1,\k_2) & = \, \frac{\k_1 \cdot \k_2}{2 k_1^2} \,.
\end{align}
The standard second-order Newtonian kernel $\mathcal{F}_2$ is generated by the non-linear gravitational evolution. $\mathcal{S}_2$ by the tidal term and the new kernel, $\mathcal{N}_2$, is generated by the convective term $\ps \cdot  \na \vp_\G$. 


\section{Three-point functions of halo and matter overdensities}\label{sec:bispectrum}

In \cref{sec:matterbispectrum,sec:halobispectrum} we will present the tree-level bispectra for the halo and matter overdensities. We adopt the definition
\begin{equation}
\label{eq:defB}
 \langle \delta^E_{\alpha}(\k_1) \delta^E_{\beta}(\k_2) \delta^E_{\gamma}(\k_3) \rangle = (2\pi)^3 \delta^D(\k_1 + \k_2 + \k_3) B_{\alpha \beta \gamma}(\k_1,\k_2,\k_3) \,,
\end{equation}
where $\alpha, \beta, \gamma = h ,m$ with the labels $h$ and $m$ standing for halo and matter respectively. 
The crossed halo-matter bispectra are given in \cref{app:crossbispetra}. 

\begin{figure}
\centering
\includegraphics[width=.45\textwidth]{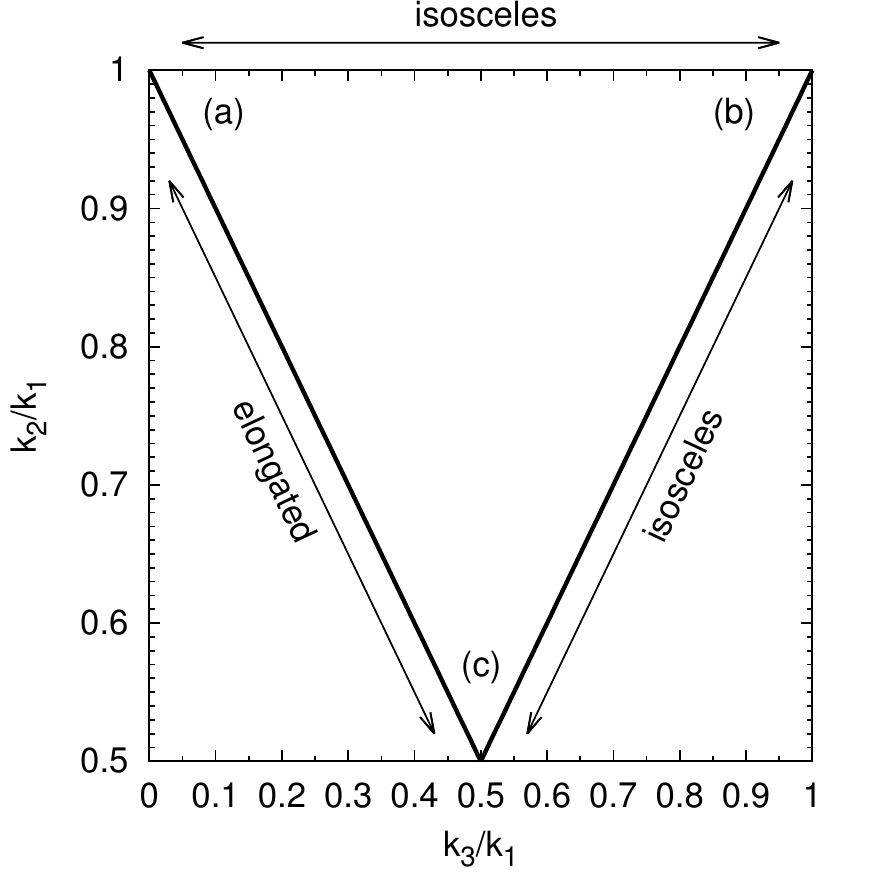}
\caption{Explanation of the visual representation for the bispectrum introduced in \cite{Jeong:2009}. The triangular-shaped region that hosts the colour map is due to the condition $k_3 \leq k_2 \leq k_1$. This requirement avoids double visualizations of the same triangular configuration. For the allowed values of $k_2/k_1$ and $k_3/k_1$ we recognise same specific configurations: point (a) is for the squeezed limit ($k_1 \simeq k_2 \gg k_3$), (b) for the equilateral configuration ($k_1=k_2=k_3$) and (c) for the folded one ($k_1 = 2 k_2 = 2 k_3$). The elongated triangles ($k_1 = k_2 + k_3$) resides on the left edge, while the upper and right edges correspond to isosceles triangles ($k_1 > k_2 = k_3$ or $k_1=k_2 > k_3$). General configurations are in the inner region.}
\label{fig:triangle}
\end{figure}

In \cref{sec:halobispectrum,sec:analysis} we will make use of the graphical representation introduced by Jeong and Komatsu \cite{Jeong:2009} to show the shape dependence of the bispectrum. The amplitude of $B_{\alpha \beta \gamma}$, or parts of it, will be plotted as a function of $k_2/k_1$ and $k_3/k_1$ in a colour map, under the condition $k_3 \leq k_2 \leq k_1$. This requirement avoids multiple visualizations of the same triangle. From all the possible choices of $k_2/k_1$ and $k_3/k_1$, we can identify some specific configurations which are shown in  \cref{fig:triangle}: equilateral ($k_1=k_2=k_3$), isosceles ($k_1 > k_2 = k_3$ or $k_1=k_2 > k_3$), folded ($k_1 = 2 k_2 = 2 k_3$), squeezed ($k_1 \simeq k_2 \gg k_3$) and elongated ($k_1 = k_2 + k_3$). 

\subsection{Matter bispectrum}\label{sec:matterbispectrum}

The matter bispectrum is given from \cref{eq:deltaEk,eq:defB}
\begin{equation}
 B_\text{mmm}(\k_1,\k_2,\k_3)=\left(2P(k_1)P(k_2)\mathcal{F}_2(\k_1,\k_2) +2\fnl\frac{P(k_1)P(k_2)\alpha(k_3)}{\alpha(k_1)\alpha(k_2)} +2\cyc\right) \,,
\end{equation}
where we have dropped the redshift $z$ dependence from the function $\alpha(k,z)$ and the matter power spectrum $P(k,z)$ in order to simplify the notation; we will do the same in the following sections.
The matter bispectrum is thus generated by both primordial non-Gaussianity, $\fnl$, and non-linear gravitational evolution, $\mathcal{F}_2$.

The tree-level approximation based on perturbation theory well describes the simulation results at scales up to $k \simeq 0.05-0.1 h \text{Mpc}^{-1}$, depending on the redshift. Including one-loop corrections can significantly extend the validity for bispectrum to $k \simeq 0.3 h \text{Mpc}^{-1}$ at redshift $z \gtrsim 1$ \cite{Sefusatti:2011}. Alternatively, it is possible to use an effective kernel $\mathcal{F}_2^\eff$ calibrated against simulations \cite{Scoccimarro:2000ee}. This phenomenological approach gives simpler expressions in the non-linear regime,  and accurate predictions for the bispectrum, up to $k \simeq 0.4 h \text{Mpc}^{-1}$ in the redshift range $0 \leq z \leq 1.5$, when Gaussian initial conditions are assumed \cite{Marin:2012}.


\subsection{Halo bispectrum}\label{sec:halobispectrum}

First let us consider the halo bispectrum when Gaussian initial conditions ($\fnl=0$) are assumed \cite{Catelan:2000vn}
\begin{align}\label{eq:gauss}
B^G_\text{hhh}(\k_1,\k_2,\k_3)=&b_{10}^3\left(2P(k_1)P(k_2) \mathcal{F}_2(\k_1,\k_2)+2\cyc\right)_\text{A} \nonumber\\
+&b_{10}^2 b_{20}\left(2 P(k_1)P(k_2)+2\cyc\right)_\text{L} \\
-&\frac{2}{7}b_{10}^2 b_{10}^\La \left(2 P(k_1)P(k_2) \mathcal{S}_2(\k_1,\k_2)+2\cyc \right)_\text{M} \nonumber \,.
\end{align}
Here and in the following the various terms appearing in the halo bispectrum are labelled in accordance with BSS \cite{Baldauf:2010vn} in order to facilitate comparison and to highlight the new terms we have identified. The halo bispectrum of \cref{eq:gauss} is generated by the non-linear gravitational evolution, $\mathcal{F}_2$, by non-linear bias, $b_{20}$, and, in particular, the last term, $\mathcal{S}_2$, is due to the non-local tidal term, $s^2$, in the expression for the halo overdensity (see \cref{eq:deltahk}). 
Note that, here and in the following, we have suppressed the E superscript in the bias factors to simplify the notation but not the L superscript. This allows to keep track of the effects generated by the non-local terms $s^2$ and $n^2$.

\begin{figure}
\centering
\includegraphics[width=1\textwidth]{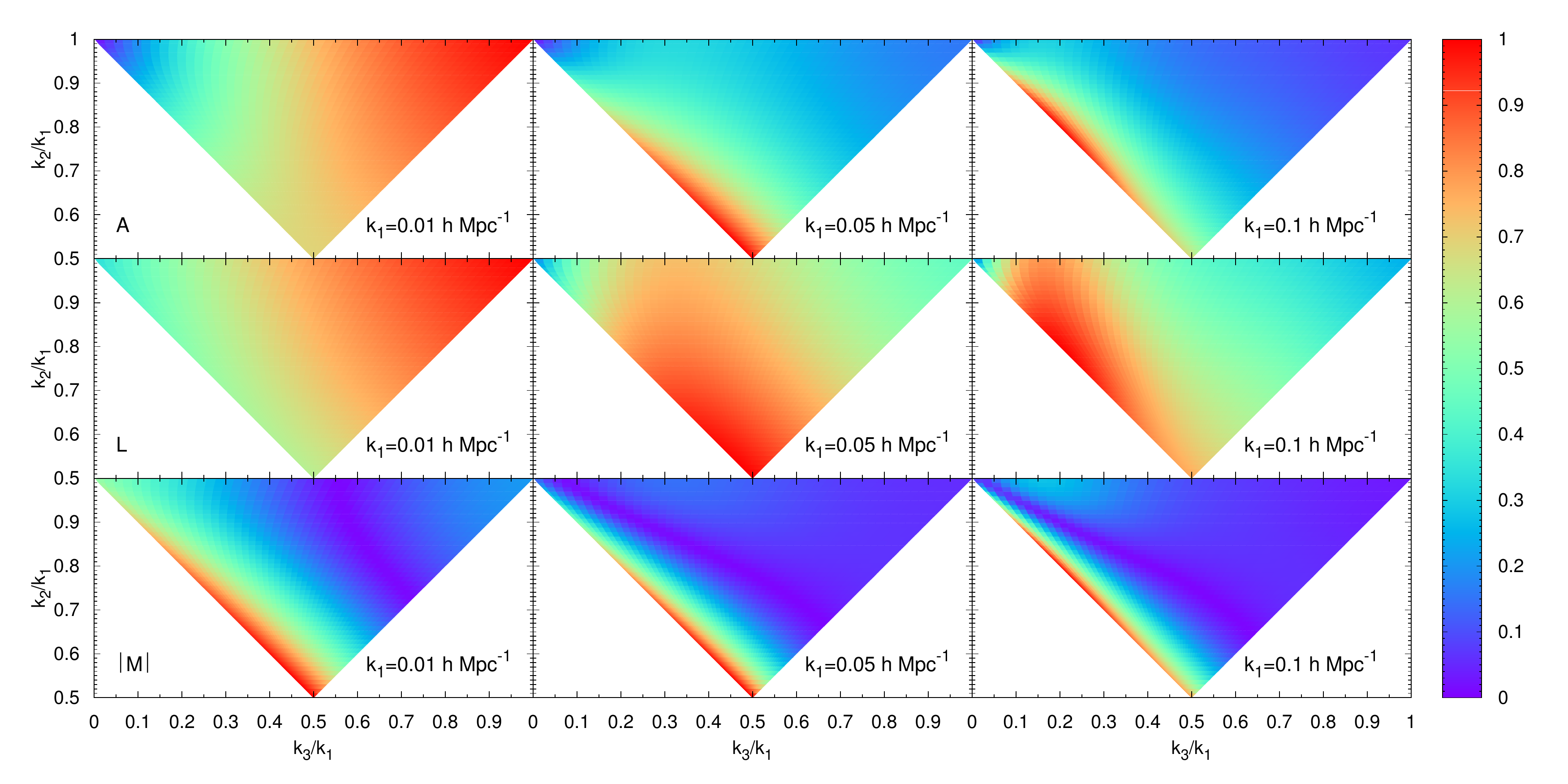}
\caption{Shape dependence of the terms contributing to the halo bispectrum when Gaussian initial conditions are assumed and for $k_1 = 0.01, 0.05, 0.1 h \text{Mpc}^{-1}$. Each term is normalized to the maximum value it can take in the $k_2/k_1$,$k_3/k_1$-space. 
This normalisation eliminates the redshift-dependence and as a result one should not make a comparison of the amplitude between plots because of the different scaling. 
 Note that the last row shows the absolute value of $M$, since it can take negative values.  The violet strip indicates where it is changing sign.}
\label{fig:gauss}
\end{figure}

In \cref{fig:gauss} we plot the shape dependence of each of the terms A,L,M for different choices of $k_1$, normalizing each of them to the maximum value it takes in the $(k_2/k_1,k_3/k_1)$-space. This normalisation eliminates the redshift-dependence and as a result one should not make a comparison of the amplitude between plots because of the different scaling. The results are shown in \cref{fig:gauss}. Note that the absolute value is plotted for M, as it can be positive or negative. Indeed, the violet region that cuts the plots into two parts is where M changes sign. 
\begin{figure}
\centering
\includegraphics[width=1\textwidth]{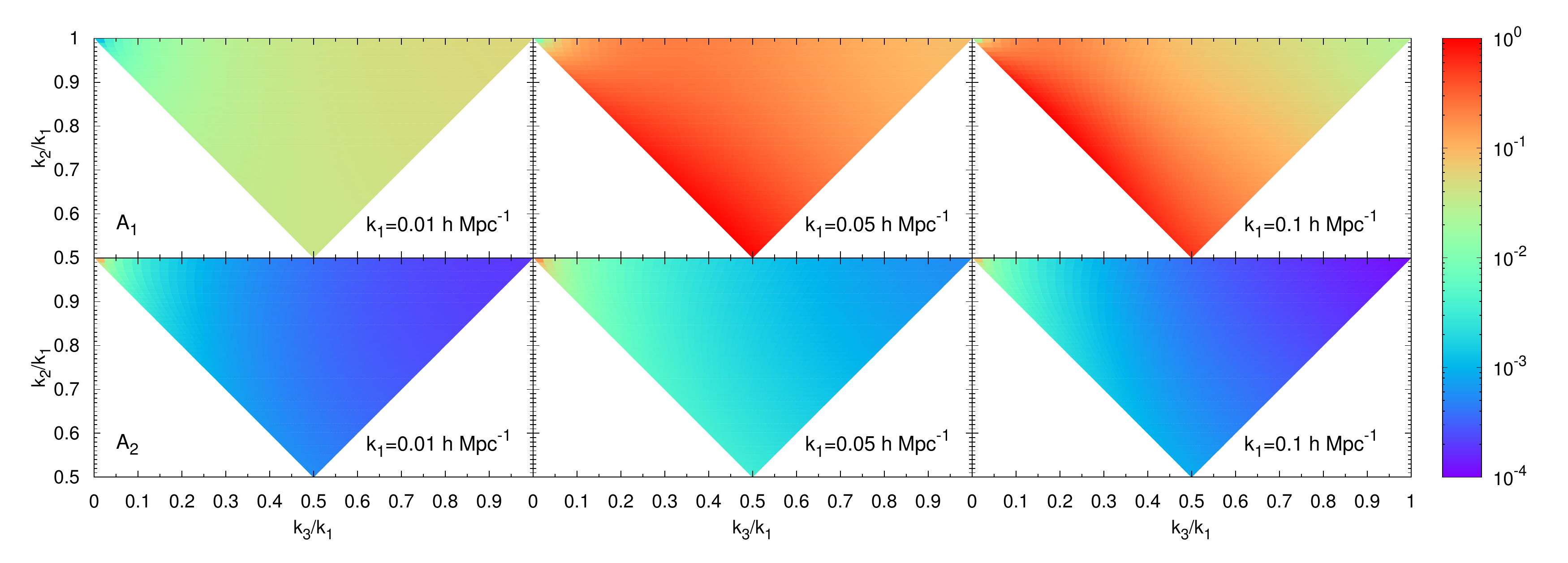}
\caption{The term A is given by the sum of two pieces, which we label $\text{A}_1$ and $\text{A}_2$. The former is sourced by non-linear gravitational evolution while the latter by PNG. We show them separately, normalized to the maximum value A can take and assuming $z=0$ and $\fnl=10$. The normalization does not completely cancel the redshift dependence, which is present in $\text{A}_2$ through the factor $D(z)^{-1}$. The effect of $\text{A}_2$ is visible in the squeezed configuration, where $\text{A}_1$ is vanishing and $\text{A}_2$ is at its maximum (the logarithmic scale might hide this aspect but a quick look back to \cref{fig:gauss} should clarify this point).}  
\label{fig:A_NG_IC}
\end{figure}

In the presence of primordial non-Gaussianity, many more terms contributes to the halo bispectrum:
\begin{align}
B_\text{hhh}(\k_1,\k_2,\k_3)=
&B_\text{hhh}^{(\text{A}\rightarrow\text{L})}(\k_1,\k_2,\k_3)\nonumber\\
-&\frac{2}{7}b_{10}^2b_{10}^\La \left(2 P(k_1)P(k_2) \mathcal{S}_2(\k_1,\k_2)+2\cyc \right)_{\text{M}}\nonumber\\ 
-&\frac{2}{7}b_{10}b_{01}b_{10}^\La \left(2 P(k_1)P(k_2)\left(\frac{1}{\alpha(k_1)}+\frac{1}{\alpha(k_2)}\right) \mathcal{S}_2(\k_1,\k_2)+2\cyc \right)_{\text{N}}\nonumber\\
-&\frac{2}{7}b_{01}^2b_{10}^\La \left(2 \frac{P(k_1)P(k_2)}{\alpha(k_1)\alpha(k_2)} \mathcal{S}_2(\k_1,\k_2)+2\cyc \right)_{\text{O}}\label{eq:bhhh}\\
-& b_{10}^2 b_{01}^\La \left( 2 P(k_1) P(k_2) \left( \frac{\mathcal{N}_2(\k_1,\k_2)}{\alpha(k_2)} + \frac{\mathcal{N}_2(\k_2,\k_1)}{\alpha(k_1)} \right) +2\cyc\right)_{\text{P}}\nonumber\\
-& b_{10} b_{01} b_{01}^\La \left( 2 P(k_1) P(k_2) \left( \frac{\mathcal{N}_2(\k_1,\k_2)}{\alpha(k_2)} + \frac{\mathcal{N}_2(\k_2,\k_1)}{\alpha(k_1)} \right)\left( \frac{1}{\alpha(k_1)} + \frac{1}{\alpha(k_2)}\right) +2\cyc\right)_{\text{Q}}\nonumber\\
-& b_{01}^2 b_{01}^\La \left( 2 \frac{P(k_1) P(k_2)}{\alpha(k_1)\alpha(k_2)} \left( \frac{\mathcal{N}_2(\k_1,\k_2)}{\alpha(k_2)} + \frac{\mathcal{N}_2(\k_2,\k_1)}{\alpha(k_1)} \right) +2\cyc\right)_{\text{R}}\nonumber \,.
\end{align}
The quantity $B_\text{hhh}^{(\text{A}\rightarrow\text{L})}$ accounts for the terms with label going from A to L; it matches exactly Eq.$(5.1)$ of BSS, that we reproduce in \cref{app:stand_halo_bispetrum}. 

The first line in \cref{eq:s_bhhh}, term A, comes from linear bias acting on the matter bispectrum. We can split this into two terms
\begin{align}
 \text{A}_1(\k_1,\k_2,\k_3) = & 2P(k_1)P(k_2) \mathcal{F}_2(\k_1,\k_2) +2\cyc \\
 \text{A}_2(k_1,k_2,k_3) = & 2\fnl\frac{P(k_1)P(k_2)\alpha(k_3)}{\alpha(k_1)\alpha(k_2)} +2\cyc \,,
\end{align}
identifying an additional term, $\text{A}_2$, with respect to the Gaussian initial conditions \cref{eq:gauss} proportional to $\fnl$. We study the shape dependence of $\text{A}_1$ and $\text{A}_2$ in \cref{fig:A_NG_IC}. Again, we plot them separately but normalize to the maximum value taken by $\text{A}(\k_1,\k_2,\k_3)=\text{A}_1+\text{A}_2$ in the $k_2/k_1$,$k_3/k_1$-space. The relative values of the two plots can thus be compared, but note that the normalization does not cancel the growth factor $1/\alpha\propto 1/D(z)$ in $\text{A}_2$ but absent in $\text{A}_1$. Hence the relative amplitude of the bispectrum from primordial non-Gaussianity $\text{A}_2$ grows with redshift relative the bispectrum  $\text{A}_1$ coming from Gaussian initial conditions.
\Cref{fig:A_NG_IC} highlights the interesting shape dependence of $\text{A}_2$; it peaks in the extremely squeezed configuration (top left), exactly where the $\text{A}_1$ term vanishes (see also \cref{fig:gauss}).

\begin{figure}
\centering
\includegraphics[width=1\textwidth]{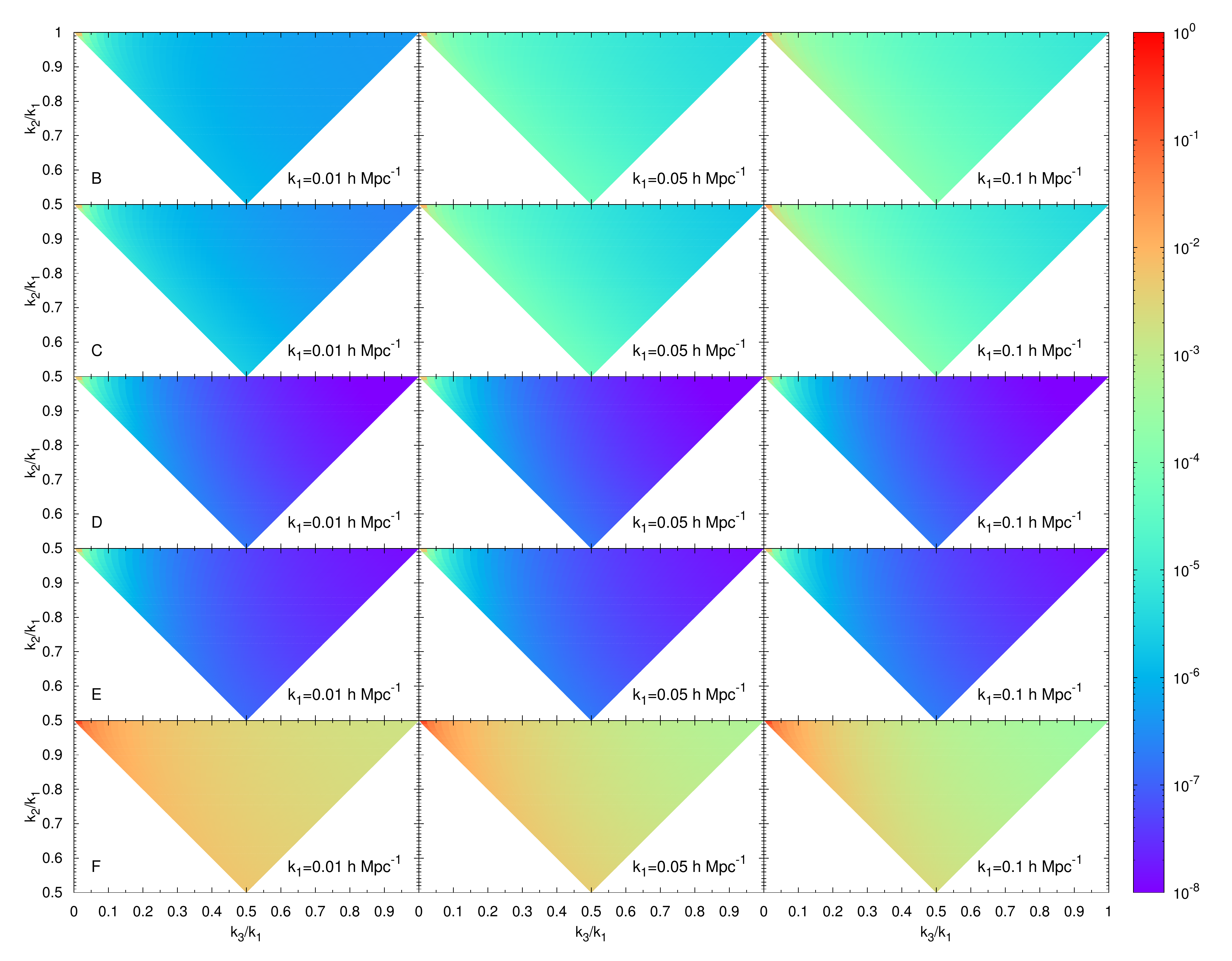}
\caption{Shape dependence of the terms, B to F, contributing to the halo bispectrum. These are generated by non-Gaussian initial conditions. A value $\fnl=10$ and redshift $z=0$ are assumed. Each term is normalized to the maximum it can take. This choice does not completely cancel the redshift dependence in the terms B and C, where it is present as a factor $D(z)^{-1}$, and does not allow a comparison of the amplitude between different plots. We clearly see that the effect of PNG is prominent in the squeezed configuration.}
\label{fig:B_F}
\end{figure}

\begin{figure}
\centering
\includegraphics[width=1\textwidth]{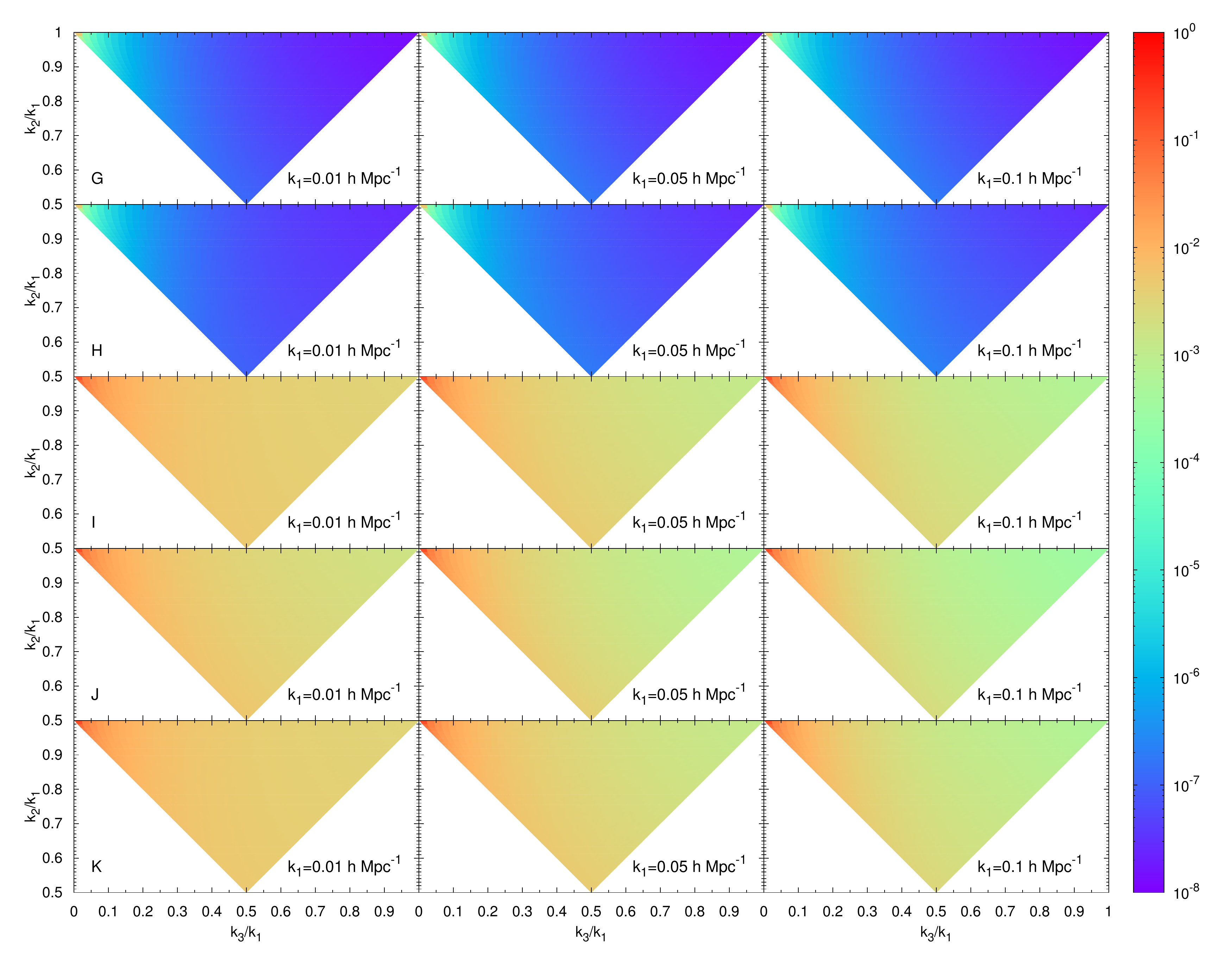}
\caption{The terms contributing to the halo bispectrum, with label going from G to K. These are generated by non-Gaussian initial conditions. A value $\fnl=10$ is assumed. Each term is normalized to the maximum it can take. This choice completely cancels the redshift dependence and does not allow a comparison of the amplitude between different terms. We clearly see that the effect of PNG is prominent in the squeezed configuration.}
\label{fig:G_K}
\end{figure}

We emphasize that $\text{A}_2$ and the terms from B to K are generated by primordial non-Gaussianity. We plot the shape dependence of the terms B to F in \cref{fig:B_F} and the term G to K in \cref{fig:G_K}. They confirm what we have previously stated: the PNG terms are greatest in the squeezed configuration.

\begin{figure}
\centering
\includegraphics[width=1\textwidth]{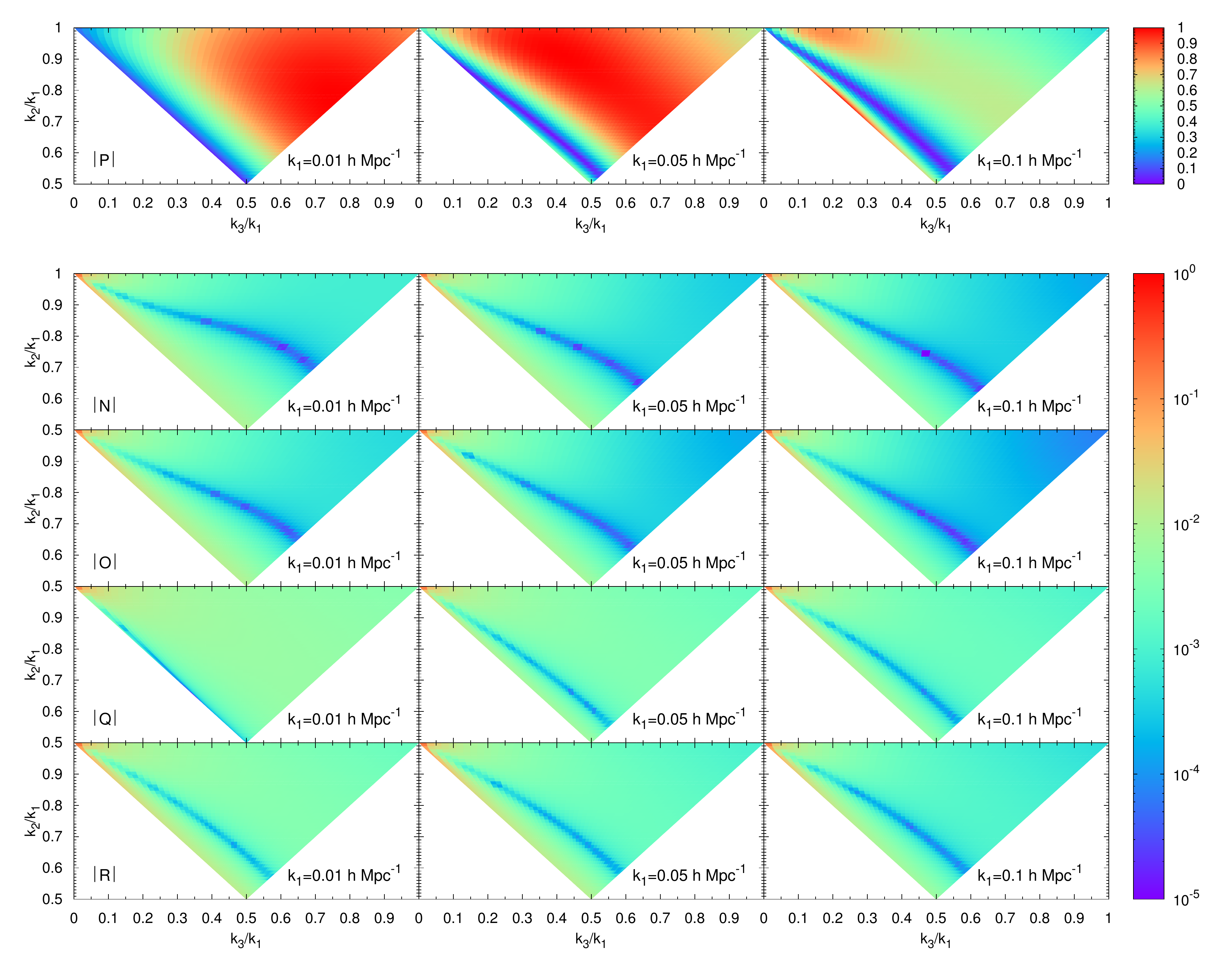}
\caption{Shape dependence of the terms with label going from N to R contributing to the halo bispectrum. These are the additional terms generated by  the presence of $s^2$ and $n^2$ in $\delta_h$. A value $\fnl=10$ is assumed. Each term is normalized to maximum value it can take so that the redshift dependence is completely dropped; note however that the different scaling does not allow one to compare the amplitude between plots. Since these terms can take negative values, we show their absolute value. Actually, the blue line indicates where they change sign, with the top right part of the plots being positive. }
\label{fig:N_R}
\end{figure}

As a result of the presence of $s^2$ and our new term $n^2$ in the expression for $\delta_h^\E$ (see \cref{eq:deltahk}), additional terms appear with respect to BSS. M, N and O that are generated by the tidal term $s^2$; schematically, they are sourced by $\langle s^2 \delta \delta\rangle$, $\langle s^2 \delta \varphi \rangle$ and $\langle s^2 \varphi \varphi \rangle$ respectively. We have seen in \cref{eq:gauss} that M is present regardless of the presence of PNG, but N and O come from the coupling between the tidal term and $\vp_\G$ for $\fnl\neq0$. On the other hand, P, Q and R are generated by $n^2$. They account for the non-local effect of the potential $\vp_\G$ and are therefore due to the presence of PNG. Schematically, they are generated by $\langle n^2 \delta \delta\rangle$, $\langle n^2 \delta \varphi \rangle$, $\langle n^2 \varphi \varphi \rangle$ respectively. In \cref{fig:N_R} we show the terms from N to R. Since they can take negative values, we plot their absolute value. Interestingly, in all of the plots we can identify a violet strip, indicating a change of sign and, hence, where the kernels $\mathcal{S}_2$ and $\mathcal{N}_2$ make each term vanish. 
\section{Analytic estimates}\label{sec:analysis}
In this section we further investigate our result for the halo bispectrum and, in particular, we compare it to our reference model BSS \cite{Baldauf:2010vn}. Since the model of BSS has shown a good fit against simulations \cite{Baldauf:2010vn,Sefusatti:2012}, we want to understand if and where differences between these two models arise. 

\begin{figure}
\centering
\includegraphics[width=1.0\textwidth]{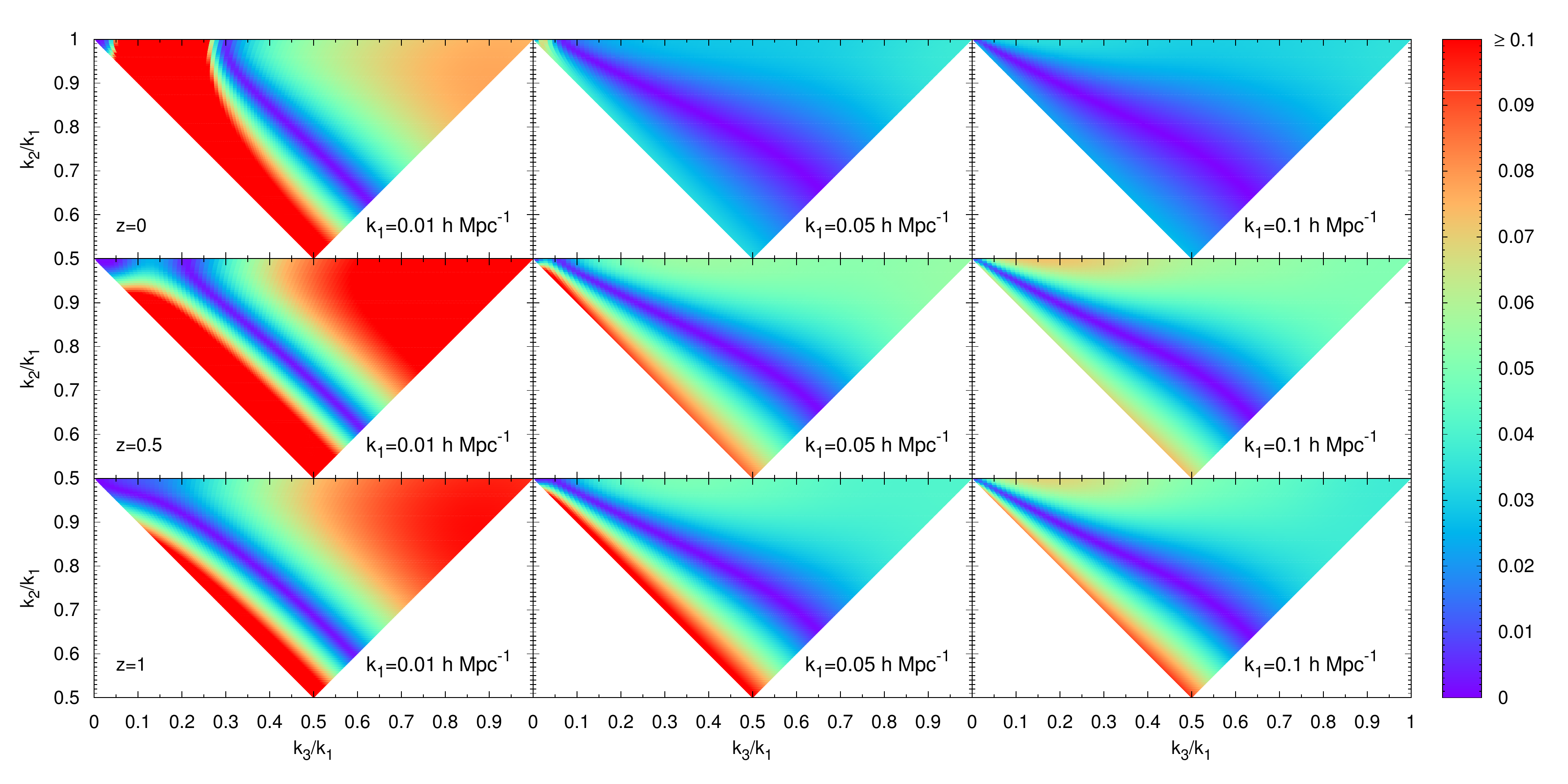}
\caption{The relative difference in absolute value between our model and the one by Baldauf et al., assuming $M=10^{13} h^{-1} M_{\odot}$ and $\fnl=10$. Note that we saturate differences above $10 \%$ to the same red colour of the palette; as the plots range between different values, this choice allows us to present them in a compact way, highlighting where the most relevant  differences are expected. However we do not observe discrepancies above $25 \%$. Interestingly, the squeezed limit is not affected while the other configurations show differences from only a few percent. A no-difference region going approximately from the squeezed to the isosceles configurations is present, following the shape dependence that we showed in \cref{fig:gauss,fig:N_R}.}
\label{fig:bss_tss}
\end{figure}

In \cref{fig:bss_tss} we plot the absolute value of the relative difference between our halo bispectrum and that of BSS
\begin{equation}
 \text{Diff}(\k_1,\k_2,\k_3) = \Biggr\vert \frac{B_\text{TRTW}(\k_1,\k_2,\k_3)-B_\text{BSS}(\k_1,\k_2,\k_3)}{B_\text{BSS}(\k_1,\k_2,\k_3)} \Biggr\vert \,,
\end{equation}
for values of $k_1 = 0.01, 0.05, 0.1 h \text{Mpc}^{-1}$ and redshift $z=0,0.5,1$. We consider halos of mass $M = 10^{13} \text{h}^{-1} M_{\odot}$ and primordial non-Gaussianity with $\fnl=10$. 
We choose to saturate differences above $10\%$ to the same red colour of the palette for the purpose of presenting many different plots with different ranges of values in a compact way. However we do not observe differences above $25\%$ \footnote{An exception is the top left plot of \cref{fig:bss_tss}, where there is a small curve close to the squeezed configuration in which the discrepancy is actually bigger than that. This because in that area the BSS halo bispectrum crosses zero for these values of the bias coefficients. }. 

For the value of the halo mass and $\fnl$ considered, we find that the terms M and P are the main sources of the differences, with all the other terms contributing very little or having a negligible effect. 
%
The most relevant differences appear for $k_1 = 0.01 h \text{Mpc}^{-1}$: up to $25 \%$ in the elongated, folded and equilateral regions for all the redshifts considered. These discrepancies drop to a few percent for $k_1 = 0.05 h \text{Mpc}^{-1}$ at $z=0$, while for $z=0.5,1$ they are reduced to about $5 \%$ when approaching the equilateral configuration and to order $10 \%$ in the elongated and folded regions. For $k_1 = 0.1 h \text{Mpc}^{-1}$ we observe a similar pattern, but at $z=0.5$ the approximately $10 \%$ difference area that was present in the elongated and folded regions for $k_1 = 0.05 h \text{Mpc}^{-1}$ is almost completely washed out, decreased to about $5 \%$. However, that area is still present for $z=1$, although reduced in size. We also recognise an area where the difference is close to zero [the  violet region going approximately from the squeezed (top left) to the isosceles (bottom right) configuration] corresponding to a vanishing contribution from the terms M to R, as shown in \cref{fig:gauss,fig:N_R}. Interestingly, in all plots the squeezed configuration is unaffected.

We can understand the features that we have just described by studying the analytic solutions of the halo bispectrum in three simple configurations: 
\begin{itemize}
 \item In the equilateral configuration, where $k_1=k_2=k_3=k$, the halo bispectrum becomes
\begin{align}
 B_\text{hhh} =&  \Biggr\{ \frac{b_{10}^2}{7} (12 b_{10} + b_{10}^\La + 42 b_{20} )  \elnn 
& + \frac{1}{\alpha(k)} \biggr[ \frac{2}{7} b_{01} b_{10} \bigr(12 b_{10} + b_{10}^\La + 42 b_{20} \bigr) + 3 b_{10}^2 \biggr( b_{01}^\La + 2 \bigr( \fnl b_{10} + b_{11} \bigr) \biggr) \biggr] 
\label{eq:equi} 
 \nonumber\\
& +\frac{1}{\alpha(k)^2} \biggr[ \frac{b_{01}^2}{7} \bigr(12 b_{10} + b_{10}^\La + 42 b_{20} \bigr) + 6 b_{01} b_{10} \biggr( b_{01}^\La + 2 \bigr( \fnl b_{10} + b_{11} \bigr) \biggr) + 6 b_{02} b_{10}^2 \biggr] \elnn
& +\frac{3}{\alpha(k)^3} \biggr[ b_{01}^2 \biggr( b_{01}^\La + 2 \bigr( \fnl b_{10} + b_{11} \bigr) \biggr) + 84 b_{01} b_{02} b_{10} \biggr] + \frac{6 b_{01}^2 b_{02}}{\alpha(k)^4} \Biggr\} P(k)^2
\end{align}
We remind the reader that the contributions of $b_{10}^\La$ and $b_{01}^\La$ indicate where the non-local, non-linear terms of our model ($s^2$ and $n^2$) are introducing differences respect to the BSS model. By looking at \cref{eq:bhhh} we notice that the terms M, N, O, P, Q, R are respectively linked to the bias combinations $b_{10}^2 b_{10}^\La$, $b_{10}b_{01} b_{10}^\La$, $b_{01}^2 b_{10}^\La$, $b_{10}^2 b_{01}^\La$, $b_{10} b_{01} b_{01}^\La$, $b_{01}^2 b_{01}^\La$, so that we can actually recognize each of them in \cref{eq:equi}; M appears in the first line of \cref{eq:equi}, while N and P appear in the second line, O and Q in the third and R in the first term on the fourth line.

The presence of a larger difference at high redshift and on large scales can be explained by noting that the function $\alpha(k)$ takes smaller values in those regimes and enhances the terms inside the square brackets (see \cref{eq:alpha} and \cref{fig:alpha&p}) and, therefore, accentuates the effect of $b_{10}^\La$ and $b_{01}^\La$.

\item In the folded configuration, $k_2= k_3 = k$ and $k_1 = 2k$, the halo bispectrum can be written as 
\begin{align}
 B_\text{hhh} = & \Biggr\{ 2 b_{10}^3 \left(2 - \Pi^2 \right) + b_{10}^2 \left(2 b_{20} - \frac{8}{21} b_{10}^\La\right) \left(1 + 4 \Pi^2 \right)+ \elnn
& + \frac{1}{\alpha(k)} \biggr[ 10 b_{10}^3 \fnl \Pi^2 + b_{10}^2 \biggr(b_{01} \left(8 - \frac{\Pi}{2} - 2 \Pi^2 \right) + 2 b_{01}^\La \left(-1 + \Pi + \Pi^2\right) + b_{11} \left(2 + \Pi + 4 \Pi^2\right)\biggr) \elnn
& +b_{10} \left( 2 b_{01} b_{20} -\frac{8}{21} b_{01} b_{10}^\La \right) \left(2 + \Pi + 4 \Pi\right) \biggr]  \elnn 
&+ \frac{1}{\alpha(k)^2}\biggr[\dots\biggr] + \frac{1}{\alpha(k)^3}\biggr[\dots\biggr] 
 + \frac{1}{\alpha(k)^4}\biggr[\dots\biggr] \Biggr\} P(k)^2 \,.
\end{align}
where we define $\Pi = T(2k)/T(k)$ to make the notation more compact. 
The full expression is long so we have given just the first two terms. These are enough to explain what we observe in \cref{fig:bss_tss}. 
Again, $b_{10}^\La$ and $b_{01}^\La$ are present. 
As for the equilateral configuration, the effect of M appears in the first term, while N and P in the second one, O and Q in the third and R in the fourth one. We see that the effect of $b_{10}^\La$ and $b_{01}^\La$ can be enhanced depending on the value of $\Pi$. When $k_1=0.01 h \text{Mpc}$ the ratio $\Pi \approx 1$, while for the other two values of $k_1$ the ratio $\Pi<1$. This suggests why bigger differences should be expected in the case with $k_1=0.01 h \text{Mpc}$. 

The same considerations apply to the function $\alpha (k)$ as in the equilateral case, so that the differences are larger at high redshift and on large scales. 

\item A simple expression for the halo bispectrum in the squeezed limit can be found by setting $k_1=k_2=\epsilon k_3=k$ with $\epsilon \gg 1$, corresponding to an isosceles triangle, whose degree of squeezing is controlled by the parameter $\epsilon$. For squeezed triangles (large values of $\epsilon$), the leading term in the bispectrum is
\begin{align}
 B_\text{hhh} \approx & \Biggr[ \frac{2}{\alpha(k)^2} \biggr(2 \fnl b_{01} b_{10}^2 + b_{01} b_{10} b_{11}\biggr) +\frac{2}{\alpha(k)^3} \biggr(2 \fnl b_{01}^2 b_{10} + b_{01}^2 b_{11} + 2 b_{01} b_{02} b_{10}\biggr) \elnn
& +\frac{4}{\alpha(k)^4} b_{01}^2 b_{02} \Biggr] P(k)^2 \epsilon^3 \,.
\end{align}
The absence of terms involving bias coefficients $b_{10}^\La$ and $b_{01}^\La$ explains why our predictions do not differ from the BSS model in extremely squeezed configurations.
\end{itemize}

\section{Conclusions}\label{sec:discussion}


 %
An important application of measurements of large-scale structure in our Universe is to determine the distribution of primordial perturbations, in particular possible non-Gaussian signatures of scenarios for the origin of structure in the very early universe. 
For example, the shape information contained in the primordial bispectrum is a valuable tool to discriminate between different inflationary models. 
The matter bispectrum at later times is due to a combination of both primordial non-Gaussianity and non-linear evolution under gravity. 
However, the way in which gravitationally collapsed halos trace the density field, the bias model, can enhance the effect of primordial non-Gaussianity in the galaxy distribution. In particular, local-type non-Gaussianity leads to a scale-dependent bias which can have a dramatic effect on very large scales in both the halo power spectrum \cite{Dalal:2007cu,Matarrese:2008nc} and the halo bispectrum \cite{Tasinato:2013vna}.

In this paper we developed a local Lagrangian bias model, focussing on second-order, non-local and non-Gaussian effects. In particular, we extended and applied a local Lagrangian biasing scheme to a general set-up with local-type primordial non-Gaussianity. For an $\fnl$ cosmology, we re-derive the known  result \cite{Giannantonio:2009ak} that the halo overdensity can be expressed as a bivariate expansion in terms of the linear matter overdensity in the Lagrangian frame and the primordial Gaussian potential. 

Non-linear evolution of the matter field in general gives rise to second-order terms in the matter density which include non-local, tidal terms (see \cref{eq:secL}), while transforming from the Lagrangian to the Eulerian frame introduces a non-local convective term at second order (see \cref{eq:secE}). Non-local here implies terms derived from derivatives of the potential, not directly from the local density or its derivatives. Both these terms are included in the usual kernel, $\mathcal{F}_2$, for the second-order density in Eulerian space. But since the halo density is determined by the linear matter overdensity in the Lagrangian frame, one must account separately for these non-local terms to reconstruct the halo density at late times. These terms are absent at early times, or if we restrict ourselves to the spherical collapse approximation (see~\cref{app:local_eul_b}).

We have shown in this paper that in the bivariate expansion we must also account at second order for the convective term relating the primordial potential in the Lagrangian frame to that in Eulerian space at later times (see \cref{eq:varphitransf}). This gives rise to a new term in the halo bispectrum in the presence of local-type primordial non-Gaussianity which has not previously been studied as far as we are aware.


Setting $\fnl=0$ and are able to recover the halo bispectrum model of \cite{Catelan:2000vn}, when a local Lagrangian biasing scheme is applied. Three terms appear in the halo bispectrum (A,L and M) that are sourced by (A) the non-linear matter density encoded in the kernel $\mathcal{F}_2$, (L) the non-linear bias $b_{20}$ and (M) the tidal term $s^2$. 

Generalising to $\fnl \neq 0$, we found $12$ terms in the halo bispectrum (labelled A to L) matching the BSS model \cite{Baldauf:2010vn}. In this case, the non-linear matter density term, A, includes a correction due to PNG, while the non-linear bias term, L, is left unchanged. The other contributions (B to K) come from a mixture of bivariate terms, involving the bias coefficients $b_{01}$, $b_{11}$, $b_{02}$. We also found a contribution, M, sourced by the tidal term, $s^2$, which also couples with terms in the halo overdensity that are specifically due to PNG and, hence, generate new contributions, N and O, in the halo bispectrum. 
The new convective term, $n^2$, also generates contributions in the presence of primordial non-Gaussianity; in the halo bispectrum we have found three new contributions, P, Q and R, due to this term.

In order to investigate the magnitude and shape of the various contributions to the bispectrum, we have implemented a version of the Sheth-Tormen mass function corrected for PNG in light of the Lo Verde mass function, following BSS \cite{Baldauf:2010vn}. This allowed us to numerically calculate the bias coefficients for our model and predict the halo bispectrum in different configurations for sample values of $\fnl$ and at various scales and redshifts.

We investigated the halo bispectrum by comparing it to the fiducial model of BSS. Assuming halos of mass $\MA = 10^{13} h^{-1} \MA_{\odot}$ and $\fnl = 10$, we  found:
  \begin{itemize}
   \item[-] At redshift $z=0$ differences up to $25\%$ in the halo bispectrum for $k_1 = 0.01 h \text{Mpc}^{-1}$ in the elongated and folded configurations and approximately $7-8\%$ when approaching the equilateral configuration, while these drop to a few percent when $k_1 = 0.05 $ or $0.1 h \text{Mpc}^{-1}$.
   \item[-] At redshift $z=0.5$ differences up to $25\%$ for $k_1 = 0.01 h \text{Mpc}^{-1}$ in the elongated, folded and equilateral configurations. When $k_1 = 0.05 h \text{Mpc}^{-1}$ differences of order $10 \%$ are still visible in the elongated and folded shapes, while they decrease to approximately $5\%$ towards the equilateral configuration.  For $k_1 = 0.1 h \text{Mpc}^{-1}$ these differences reduce to about $5\%$.
  \item[-] At redshift $z=1$ we find results similar to 
  those for $z=0.5$, except that 
  differences of order $10 \%$ are still visible in the elongated and folded regions for $k_1 = 0.1 h \text{Mpc}^{-1}$.
  \end{itemize} 
In general, we observe that the non-local terms have a negligible effect in extremely squeezed configurations.

Our results indicate that the non-local terms in the halo overdensity could have a significant contribution to the galaxy bispectrum, especially on large scales and at high redshift. The next challenge is to test these theoretical predictions against N-body simulations with non-Gaussian initial conditions. 
Ultimately we would wish to be able to estimate the signal-to-noise for upcoming surveys, like the ESA Euclid mission \cite{EUCLID}, which probes large scales and high redshifts, in order to explore the observability of these non-local effects in the bispectrum.
A full discussion of the observability of these effects must include a halo occupation model, to describe how galaxies populate halos and many other effects including redshift space distortions and lensing along the line of sight, from galaxies to the observer, in order to translate theoretical model for the halo bispectrum into predictions for the observed galaxy angular bispectrum in redshift space.
Nonetheless, we have identified in this paper novel contributions to the expected galaxy bispectrum on large scales and high redshift in the presence of primordial non-Gaussianity.

\acknowledgments
The authors wish to thank Cornelius Rampf for help in the early stages of this work and Tommaso Giannantonio, Nina Roth, Jennifer E. Pollack for many helpful discussions. They are also grateful to Kazuya Koyama for carefully reading the paper and providing many useful comments. MT acknowledges Tobias Baldauf for email communication and Vincent Desjacques, H\'ector Gil-Mar\'in, Francesco Pace and Sandro Ciarlariello for useful discussions. GT is supported by an STFC Advanced Fellowship ST/H005498/1. DW is supported by STFC grants ST/K00090X/1 and ST/L005573/1.

\appendix

\section{Displaced Gaussian random fields}\label{sec:displaced}

In deriving the halo abundance in Eulerian space we need to map the initial gravitational potential in Lagrangian coordinates, \cref{eq:Phi}, into Eulerian coordinates under the coordinate displacement \cref{eq:lag_to_eu}, which is itself determined by the gravitational potential. In this appendix we shall demonstrate how even an initial Gaussian field in Lagrangian coordinates may be transformed into a non-Gaussian field by this displacement to Eulerian coordinates.

We start from a random field, $\hat\varphi_G(q)$, defined with respect to the Lagrangian coordinate chart, $q$. Let
\begin{equation}
\hat\varphi_G(q) = \epsilon f(q) \hat{a} \,,
\end{equation}
where $\hat{a}$ denotes a Gaussian random variable and $\epsilon$ is a small perturbative parameter. Thus $\hat\varphi_G(q)$ is a Gaussian random field (first order with respect to $\epsilon$) with, for example, vanishing 3-point function
\begin{equation}
\langle \hat\varphi_G(q_1) \hat\varphi_G(q_2) \hat\varphi_G(q_3) \rangle = \epsilon^3 f(q_1) f(q_2) f(q_3) \langle \hat{a}^3 \rangle = 0 \,.
\end{equation}
where angle-brackets denote the ensemble average.

Let the Eulerian coordinate $x$ be related to $q$ by a first-order displacement field $\hat\psi(q)$, correlated with the field $\hat\varphi_G$ such that
\begin{equation}
\hat{x}(q) = q + \epsilon \psi(q) \hat{a} \,.
\end{equation}
If we consider a fixed coordinate $q$ then $\hat{x}(q)$ is itself a random variable, correlated with $\hat\varphi_G(q)$. We can then construct the field
\begin{equation}
\hat\varphi_G(\hat{x}(q)) = \hat\varphi_G(q) + \epsilon^2 \psi(q) f'(q) \hat{a}^2 + {\mathcal O}(\epsilon^3) \,,
\end{equation}
which is Gaussian at first order in $\epsilon$, but non-Gaussian at second order.
For example, the 3-point function of $\hat\varphi(\hat{x}(q))$ with respect to the coordinate chart $q$ is non-vanishing at fourth order
\[
\langle \hat\varphi_G(\hat{x}(q_1)) \hat\varphi_G(\hat{x}(q_2)) \hat\varphi_G(\hat{x}(q_3)) \rangle_q 
= \epsilon^4 \left[ \psi(q_1)f'(q_1)f(q_2)f(q_3) + {\rm perms} \right] \langle \hat{a}^4 \rangle + {\mathcal O}(\epsilon^6) \neq 0
\]

Conversely, if we work with respect to the Eulerian coordinate chart $x$, it is the Lagrangian coordinate that becomes a random field at fixed coordinate $x$:
\begin{equation}
\hat{q}(x) = x - \epsilon \psi(x) \hat{a} + {\mathcal O}(\epsilon^2) \,,
\end{equation}
and $\hat\varphi_G(\hat{q}(x))$ becomes a non-Gaussian field at second order
\begin{equation}
\hat\varphi_G(\hat{q}(x)) = \hat\varphi_G(x) - \epsilon^2 \psi(x) f'(x) \hat{a}^2 + {\mathcal O}(\epsilon^3) \,,
\end{equation}
where $\hat\varphi_G(x) = \epsilon f(x) \hat{a}$ is a first-order Gaussian random field in Eulerian space.
 
\section{Halo mass function}\label{app:massfunction}

The Press-Schechter formalism \cite{1974ApJ...187..425P} and its extensions \cite{Bardeen:1986,Bond:1991} provide a model to describe the full non-linearly evolved density field, and in particular the
number of gravitationally collapsed dark matter halos, in terms of the initial, linearly growing density field (see \cite{Zentner:2006vw} for a pedagogical review).
Dark matter halos are identified as peaks in the linearly growing density field of \cref{eq:alpha}, exceeding a suitable threshold value, $\delta_c$. This is usually assumed to be the linearly growing density mode for a spherically collapsed object. For example, in a flat $\Lambda$CDM universe the threshold for a spherical collapsed halo is \cite{Kitayama:1996ne}
\begin{equation}
 \delta_c (z) = \frac{3 (2\pi)^{2/3}}{20} \left[ 1 + 0.0123 \log \Omega_m (z)\right] \, ,
\end{equation}
reducing to $\delta_c \simeq 1.686$ when $\Omega_m = 1$. Hence $\delta_c$ is  weakly dependent on the value of $\Omega_m$ and $\Omega_{\Lambda}$ at the time of collapse \cite{Eke:1996ds}. 

For Gaussian initial conditions, the smoothed first-order density field is a Gaussian field with variance
\begin{equation}
    \sigma_\G^2 (\MA,z) = \frac{D^2 (z)}{2 \pi^2} \int dk \, k^2 \, W_\MA^2(k,R) P_0 (k) \,,
\end{equation}
where $P_0 (k)$ is the linear matter power spectrum at redshift $z=0$ (see \cref{fig:alpha&p}) and $W_\MA (k,\R)$ is a window function in Fourier space that filters out modes below the length scale $\R(\MA)=(3\MA/4 \pi \rho_m)^{1/3}$. We will adopt the real-space top-hat filter, with Fourier transform
\begin{equation}
 W_\MA (k,\R) = \frac{3}{(k \R)^3} \left[ \sin (k \R) - k \R \cos(k \R) \right]\, .
\end{equation}

The Press-Schechter (PS) approach predicts the number density of objects with mass $\MA$ at redshift $z$, i.e. the \emph{mass function}, to be
\begin{equation}
 \label{eq:nu}
  n_h (\MA,z) = f(\nu) \frac{\rho_m}{\MA} \biggr\vert \frac{d \ln \sigma_\G}{d \MA} \biggr\vert \,,
\end{equation}
where we  introduce  the variable $\nu = \delta_c / \sigma_\G $. The analytic form for the PS distribution is
\begin{equation}\label{eq:ps}
  f_\PS (\nu) = \sqrt{\frac{2}{\pi}} \nu e^{-\frac{\nu^2}{2}} \,.
\end{equation}

By considering the collapse of ellipsoidal overdense regions, the Sheth-Tormen (ST) distribution \cite{Sheth:1999,Sheth:2001,Sheth:2002} is obtained using
\begin{equation}\label{eq:st}
  f_\ST (\nu) = A(p) \sqrt{\frac{2 \gamma}{\pi}} \left[ 1 + \left( \gamma \nu^2 \right)^{-p}\right] \nu e^{-\gamma \, \frac{\nu^2}{2}} \,,
\end{equation}
where the parameters $\gamma = 0.707$ and $p = 0.3$ are found by fitting against simulations and $A(p) = 0.322$ under the requirement that all the mass is collapsed into halos. \Cref{eq:st} greatly improved the agreement with simulations.
Interestingly, both the PS and ST functions, $f$, depend only on the variable $\nu$, for this reason they are know as \emph{universal} mass functions.

In presence of weakly non-Gaussian initial conditions, the probability distribution function (PDF) of fluctuations can be approximated by an Edgeworth expansion. Then, a derivation similar to the one yielding to \cref{eq:ps} leads to \cite{LoVerde:2007ri}
\begin{equation}\label{eq:lv}
  f_\LV(\nu,\MA) = f_\PS(\nu) \left[ 1 + \frac{1}{6}\left(\kt(\MA) H_3(\nu) - \frac{d\kt(\MA)/d\MA}{d\ln\sigma^{-1}/d\MA} \frac{H_2(\nu)}{\nu} \right) \right] \,.
\end{equation}
$H_n$ is the $n$-th Hermite polynomial and $\kt(\MA)$ the $3$rd cumulant, defined as $\kt(\MA) = \langle \delta_\MA^3 \rangle / \sigma^3$, where
\begin{equation}
 \langle \delta_\lin^3 \rangle = \int \dqc \int \dqcp \int \dqcpp W_\MA(p) \alpha(p,z) W_\MA(p') \alpha(p',z) W_\MA(p'') \alpha(p'',z) \langle \Phi_\ini(\p) \Phi_\ini(\p') \Phi_\ini(\p'') \rangle \,.
\end{equation}
In addition, for primordial non-Gaussianity of the form of \cref{eq:png}, the variance needs to be replaced with
\begin{equation}\label{eq:kd}
  \begin{split}
      \sigma^2 = \langle \delta_\lin^2 \rangle & = \int \dqc \int \dqcp W_\MA(p) \alpha(p,z) W_\MA(p') \alpha(p',z) \langle \Phi_\ini(\p) \Phi_\ini(\p') \rangle \\
					    & \approx \sigma_G^2 \left( 1 + \kd(\MA) \right) \, ,
  \end{split}
\end{equation}
\Cref{eq:lv} is known as the Lo Verde et al. (LV) mass function. In \cite{2011JCAP...08..003L} fitting functions for $\kd$ and $\kt$ are given; although $\kd \propto \fnl^2$, it gives a negligible correction to $\sigma_G$ for any realistic value of $\fnl$ and we will neglect it here\footnote{In general $\kd \propto \tnl^2 / \fnl^2$. The model of \cref{eq:png} satisfies the Suyama-Yamaguchi equality: $\tnl=(6/5)^2 \fnl^2$ \cite{Suyama:2007bg}.}. The $3$rd cumulant $\kt$ reads  
\begin{align}
  \label{eq:kt}
  \kt (\MA) & \approx \, \fnl \left( 6.6 \times 10^{-4} \right) \left[ 1 - 0.016 \ln \left( \frac{\MA}{h^{-1} \MA_{\odot}} \right) \right] \, ;
\end{align}
we refer the reader to \cref{app:a} for further details on $\kt$ and $\frac{d\kt(\MA)/d\MA}{d\ln\sigma^{-1}/d\MA}$.

\begin{figure}
\centering
\subfloat
{\includegraphics[width=.45\textwidth]{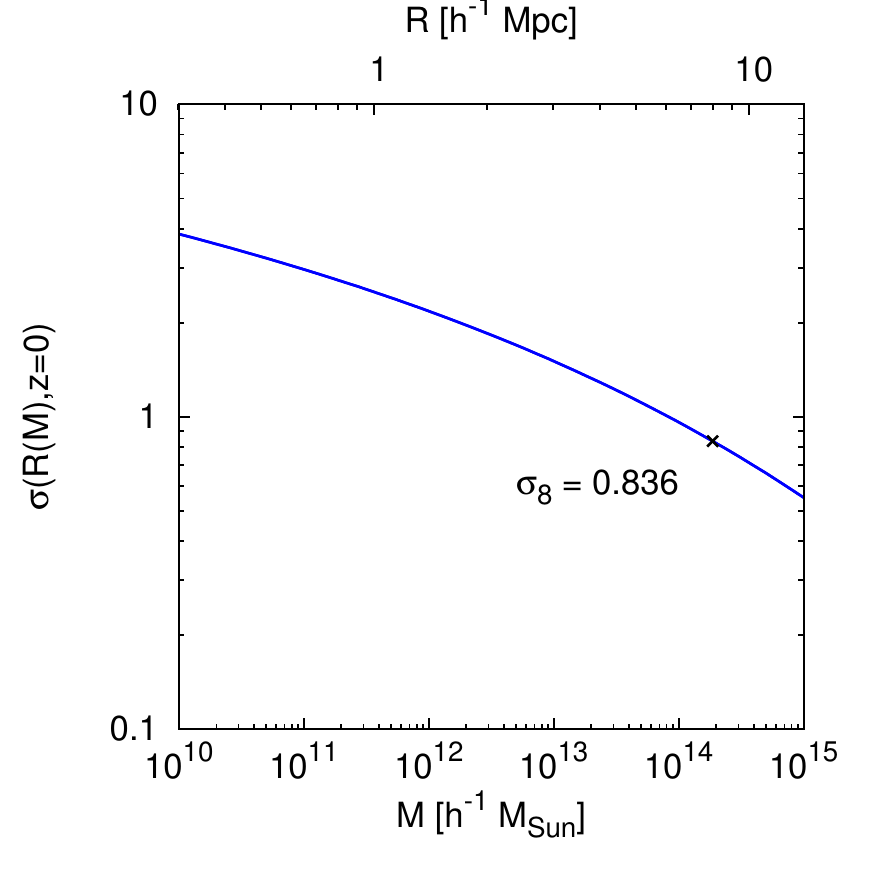}} \quad
\subfloat
{\includegraphics[width=.45\textwidth]{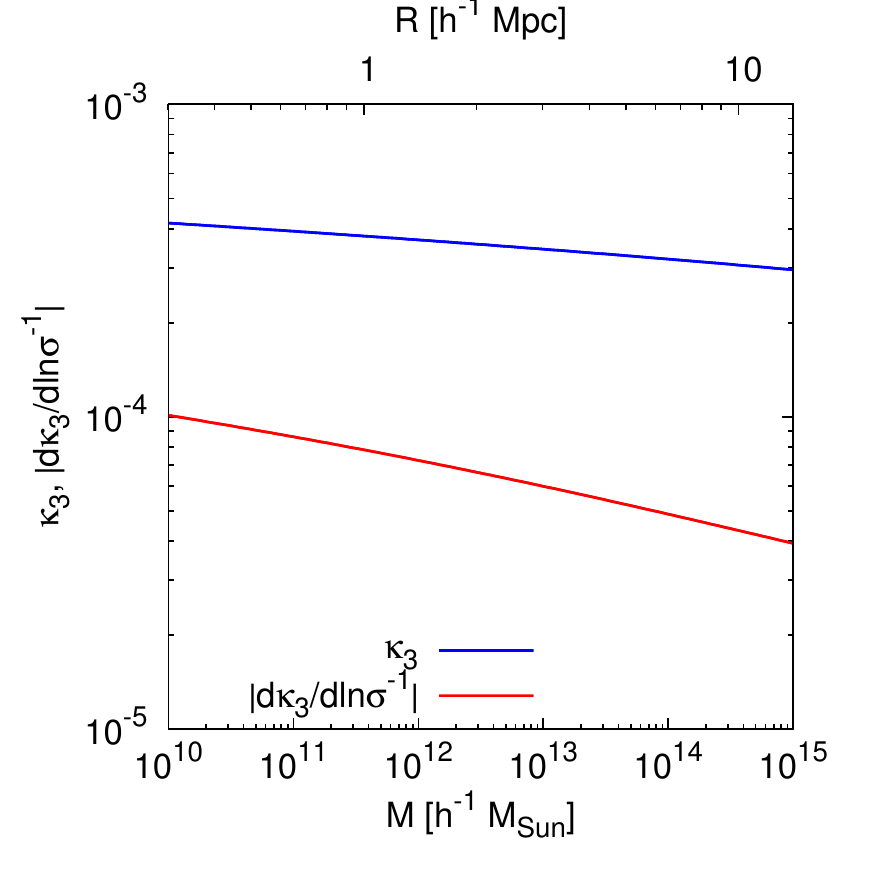}}
\caption{The left panel shows the variance $\sigma$ at redshift $z=0$. Note that values of $\fnl$ inside the current constraints from CMB and LSS produce no appreciable effects on $\sigma$. On the right side the $2$nd and $3$rd cumulant and the derivative of the latter are shown, assuming $\fnl=1$.}
\label{fig:sigma&kappa3}
\end{figure}

Note that the LV mass function is no longer universal, although the explicit dependence on the mass $\MA$ through \cref{eq:kd,eq:kt} is weak (see \cref{fig:sigma&kappa3}).
Throughout the paper we follow BSS and use an effective form of the LV mass function by replacing $f_{LV}$ with
\begin{equation}\label{eq:stlv}
 f_\LV \longrightarrow f_\LV \, \frac{f_\ST}{f_\PS} \,;
\end{equation}
the resulting mass function can be thought as the ST mass function corrected for non-Gaussian initial conditions.

\section{Comment on the fitting functions} \label{app:a}
In \cite{Smith:2011ub}, useful fitting functions for $\kt$ and $\dkt$ are given
\begin{align}
 \ktl = & \, 0.000329 \left(1 + 0.09z \right) b_1^{-0.09} \elnn
 \dktl = & -0.000061 \left(1 + 0.22z \right) b_1^{-0.25} \nonumber
\end{align}
where $b_1= (\nu^2-1)/\delta_c + 1$ is the Eulerian Gaussian bias derived from the PS mass function with $\delta_c = 1.42$. 

As we use a combination of the ST and LV mass function with $\delta_c = 1.686$, we prefer to use \cref{eq:kt} for $\kt$ and get $\dkt$ out of this one. We obtain
\begin{equation}
 \dkt = \frac{1.056 \times 10^{-5} \fnl}{\frac{\tau_G^2}{\sigma_G^2} -\frac{4.2 \times 10^{-10}}{1+\kd} \fnl^2 -1} \,,
\end{equation}
where we introduced the quantity $\tau_G^2$ that is defined as
\begin{equation}
 \tau_G^2 \equiv \int \dk W_M j_0(kR) P_0(k) \,,
\end{equation}
and $j_0(kR)$ is the spherical Bessel function of order $0$. $\kt$ and $\dkt$ are plotted in \cref{fig:sigma&kappa3}.

\section{Coordinate Jacobian}\label{app:transf} 
Since the matter density is a $3$-scalar, we have
\begin{equation}\label{eq:lagr_to_euler}
M = \rho(\x,z)  a(z)^3 \sqrt{|g_x|} d^3\x = \rho(\q,z) a(z)^3 \sqrt{|g_q|} d^3\q \,,
\end{equation}
the determinant of the Eulerian metric $g_x$ is simply $|g_x|=1$, but the Lagrangian space has a non-trivial metric $g_q$ even in Newtonian theory. If we define the coordinate Jacobian
\begin{equation}
{\cal J} \equiv  \left| \frac{d^3\x}{d^3\q} \right| = \sqrt{|g_q|} \,,
\end{equation}
by using \cref{eq:lagr_to_euler} we have to first order
\begin{equation}
{\cal J} \simeq 1 + \nabla\cdot\ps \,.
\end{equation}
At the initial redshift $z_\ini$, the Eulerian and Lagrangian frame are equivalent, i.e. $\ps=0$, and hence ${\cal J}=1$. If in addition we assume that the mass per volume element is conserved and the initial density was uniform, $\rho(\q,z_\ini)=\bar\rho(z_\ini)$ in the limit $z_\ini \to \infty$, we have
\begin{equation}
M = a^3(z) \rho(\x,z) d^3\x = a^3(z_\ini) \bar\rho(z_\ini) d^3\q
\end{equation}
and hence
\begin{equation}
\label{eq:J}
{\cal J} \equiv  \left| \frac{d^3\x}{d^3\q} \right| = \frac{a^3(z_\ini) \bar\rho(z_\ini)}{a^3(z) \rho(\x,z)} =  \frac{\bar\rho(z)}{\rho(\x,z)} = [1+\delta^\E (\x,z)]^{-1} \,,
\end{equation}
where $\delta^\E (\x)$ is the Eulerian fully non-linear density contrast.

\section{Spherical collapse approximation: Local Eulerian biasing}\label{app:local_eul_b} 
The spherical collapse approximation has been used in many works (see for instance \cite{Mo:1995cs,Giannantonio:2009ak,Baldauf:2010vn}) and we present it here with the aim of highlighting the differences respect to the more general result we derived in \cref{sec:eulerian}.

The spherical collapse corresponds to the special case in which $\ps=0$ and the velocity field vanishes at the centre of the symmetrical collapse so that the Eulerian and Lagrangian coordinates coincide at all times, $\x \equiv \q$. The results that we are going to quote all relate the density at the centre of collapse or assume a uniform density equal to that at the centre, i.e. edge effects are not considered.

Following \cite{Mo:1995cs} (see also \cite{Matsubara:2011PhRvD}), the linearly growing density at the centre of the collapse relates to the non-linear density field
\begin{equation}\label{eq:sc.delta}
  \delta_\lin^\La(\q) = \sum_{i=0}^{\infty} a_i \delta (\x)^i = a_1 \delta + a_2 \delta^2 + \dots \,, 
\end{equation}
where the coefficients are
\begin{equation} \label{eq:coefficients}
  a_1 = 1, \quad a_2 = -\frac{17}{21} \,. 
\end{equation}
\Cref{eq:sc.delta} is local in either Eulerian or Lagrangian coordinates since they coincide at the centre of the symmetrical collapse. 

On the other hand, at the centre of collapse the auxiliary potential is simply
\begin{equation}\label{eq:vp_q_x}
  \vp_\G (\q) \equiv \vp_\G (\x) \,,
\end{equation}
where we remember that $\vp_\G (\x)$ always refer to the primordial value as it simply allows to keep track of the PNG effects in the initial density field; hence $\vp_\G (\x)$ does not evolve with time like a physical quantity do and its coordinate transformation is straightforward.

Now, using \cref{eq:sc.delta,eq:vp_q_x} to express $\delta_h^\La (\q) = \delta_h^\La (\q(\x,\tau))$, we can solve \cref{eq:nhE} for $\delta_h^\E (\x)$. By using the same bias redefinition of \cref{eq:eul_bias}, the Eulerian halo overdensity finally reads
\begin{equation}
\begin{split}\label{eq:sc.deltahx}
 \delta_h(\x)=\,& b_{10}^\E \delta + b_{01}^\E \vp_\G + b_{20}^\E \delta^2 + b_{11}^\E \delta \vp_\G + b_{02}^\E \vp^2_\G \,.
\end{split}
\end{equation}
We see that the local Lagrangian biasing of \cref{eq:bivariate} is compatible with the local Eulerian model we have just obtained, despite the fact that the transformation of \cref{eq:nhE} is inherently non local: this is true only when spherical collapse dynamics applies.

\section{BSS halo bispectrum}\label{app:stand_halo_bispetrum}
The halo bispectrum model of BSS \cite{Baldauf:2010vn} predicts
\begin{align}
B_\text{hhh}^{(\text{A}\rightarrow\text{L})}(\k_1,\k_2,\k_3)=&b_{10}^3\left(2P(k_1)P(k_2) \mathcal{F}_2(\k_1,\k_2) +2\fnl\frac{P(k_1)P(k_2)\alpha(k_3)}{\alpha(k_1)\alpha(k_2)} +2\cyc\right)_\text{A}\nonumber\\
+&b_{10}^2b_{01}\left(2P(k_1)P(k_2)\left(\frac{1}{\alpha(k_1)}+\frac{1}{\alpha(k_2)}\right)F_2(\vec k_1,\vec k_2)\right.\nonumber\\
+&\left.2\fnl\frac{P(k_1)P(k_2)\alpha(k_3)}{\alpha(k_1)\alpha(k_2)}\left(\frac{1}{\alpha(k_1)}+\frac{1}{\alpha(k_2)}\right)+2\cyc\right)_\text{B}\nonumber\\
+&b_{10}b_{01}^2\left(2\frac{P(k_1)P(k_2)}{\alpha(k_1)\alpha(k_2)} \mathcal{F}_2(\k_1,\k_2)+2\fnl\frac{P(k_1)P(k_2)\alpha(k_3)}{\alpha^2(k_1)\alpha^2(k_2)}+2\cyc\right)_\text{C}\nonumber\\
+&b_{01}^2b_{02}\left(2\frac{P(k_1)P(k_2)}{\alpha^2(k_1)\alpha^2(k_2)}+2\cyc\right)_\text{D}\nonumber\\
+&b_{01}b_{10}b_{02}\left(2\frac{P(k_1)P(k_2)}{\alpha(k_1)\alpha(k_2)}\left(\frac{1}{\alpha(k_1)}+\frac{1}{\alpha(k_2)}\right)+2\cyc\right)_\text{E}\nonumber\\
+&b_{10}^2b_{02}\left(2\frac{P(k_1)P(k_2)}{\alpha(k_1)\alpha(k_2)}+2\cyc\right)_\text{F}\label{eq:s_bhhh}\\
+&b_{01}^2b_{11}\left(\frac{P(k_1)P(k_2)}{\alpha(k_1)\alpha(k_2)}\left(\frac{1}{\alpha(k_1)}+\frac{1}{\alpha(k_2)}\right)+2\cyc\right)_\text{G}\nonumber\\
+&b_{01}b_{10}b_{11}\left(P(k_1)P(k_2)\left(\frac{1}{\alpha^2(k_1)}+\frac{2}{\alpha(k_1)\alpha(k_2)}+\frac{1}{\alpha^2(k_2)}\right)+2\cyc\right)_\text{H}\nonumber\\
+&b_{10}^2b_{11}\left(P(k_1)P(k_2)\left(\frac{1}{\alpha(k_1)}+\frac{1}{\alpha(k_2)}\right)+2\cyc\right)_\text{I}\nonumber\\
+&b_{01}^2b_{20}\left(2\frac{P(k_1)P(k_2)}{\alpha(k_1)\alpha(k_2)}+2\cyc\right)_\text{J}\nonumber\\
+&b_{01}b_{10}b_{20}\left(2P(k_1)P(k_2)\left(\frac{1}{\alpha(k_1)}+\frac{1}{\alpha(k_2)}\right)+2\cyc\right)_\text{K}\nonumber\\
+&b_{10}^2b_{20}\left(2 P(k_1)P(k_2)+2\cyc\right)_\text{L}\nonumber \,.
\end{align}
\Cref{eq:s_bhhh} is incorporated in our prediction for the halo bispectrum but new terms appear (see \cref{eq:bhhh}).

\section{Halo-matter bispectra}\label{app:crossbispetra}
Here we consider the case of crossed bispectra between halos and matter. Potentially, weak lensing measurements will allow to cross correlate the dark matter density field with galaxies in the future. Also, these results provide additional predictions of our model that can be tested against simulations.  

We start by quoting the halo-halo-matter bispectrum in presence of Gaussian initial conditions
\begin{align}
B_\text{hhm}(\k_1,\k_2,\k_3)=&b_{10}^2 \biggl(6 \mathcal{F}_2(\k_1,\k_2)P(k_1)P(k_2)+2\cyc\biggr)_{\text{A}} \elnn
+&b_{10}b_{20}\biggl(4P(k_1)P(k_2)+2\cyc\biggr)_{\text{D}} \label{eq:bhhmg}\\
-&\frac{2}{7}b_{10}b_{10}^\La\biggl(2 \mathcal{S}_2(\k_1,\k_2)P(k_1)P(k_2)+2\cyc\biggr)_{\text{J}} \nonumber \,.
\end{align}
Again, we recognise that it is sourced by non-linear gravitational evolution and non-linear bias, while the last term is generated by $s^2$ in \cref{eq:deltahk}.

Assuming PNG, the halo-halo-matter bispectrum\footnote{Note that the term H corrects a typo present in Eq.$(5.6)$ of BSS.} reads
\begin{align}
B_\text{hhm}(\k_1,\k_2,\k_3)=&b_{10}^2 \biggl(6 \mathcal{F}_2(\k_1,\k_2)P(k_1)P(k_2)+6\fnl \alpha(k_3) \frac{P(k_1)}{\alpha(k_1)}\frac{P(k_2)}{\alpha(k_2)} +2\cyc\biggr)_{\text{A}} \elnn
+&  b_{01} b_{10} \biggl( 4 \mathcal{F}_2(\k_1,\k_2)P(k_1)P(k_2) \left(\frac{1}{\alpha(k_1)}+\frac{1}{\alpha(k_2)}\right) \elnn
+& 4  \fnl \alpha(k_3)P(k_1)P(k_2) \left(\frac{1}{\alpha^2(k_1)\alpha(k_2)}+\frac{1}{\alpha(k_1)\alpha^2(k_2)} \right)+2\cyc\biggr)_{\text{B}} \elnn
+&b_{01}^2\biggl(2 \mathcal{F}_2(\k_1,\k_2)\frac{P(k_1)P(k_2)}{\alpha(k_1)\alpha(k_2)} +2 \fnl\alpha(k_3)\frac{P(k_1)P(k_2)}{\alpha^2(k_1)\alpha^2(k_2)}+2\cyc\biggr)_{\text{C}} \elnn
+&b_{10}b_{20}\biggl(4P(k_1)P(k_2)+2\cyc\biggr)_{\text{D}} \label{eq:bhhm}\\
+&b_{20}b_{01}\biggl(2P(k_1)P(k_2)\left(\frac{1}{\alpha(k_1)}+\frac{1}{ \alpha(k_2)} \right)+2\cyc\biggr)_{\text{E}} \elnn
+&b_{11}b_{10}\biggl(2P(k_1)P(k_2)\left(\frac{1}{\alpha(k_1)}+\frac{1}{\alpha(k_2)}\right)+2\cyc \biggr)_{\text{F}} \elnn
+&b_{11}b_{01}\biggl(P(k_1)P(k_2)\left(\frac{1}{\alpha^2(k_1)}+\frac{2}{\alpha(k_1)\alpha(k_2)}+\frac{1}{\alpha^2(k_2)} \right) +2\cyc \biggr)_{\text{G}} \elnn
+&b_{10}b_{02}\biggl(4\frac{P(k_1)P(k_2)}{\alpha(k_1)\alpha(k_2)}+2\cyc\biggr)_{\text{H}} \elnn
+&b_{02}b_{01}\biggl(2P(k_1)P(k_2)\left(\frac{1}{\alpha^2(k_1)\alpha(k_2)}+\frac{1}{\alpha(k_1)\alpha^2(k_2)}\right)+2\cyc\biggr)_{\text{I}} \elnn
-&\frac{2}{7}b_{10}b_{10}^\La\biggl(2 \mathcal{S}_2(\k_1,\k_2)P(k_1)P(k_2)+2\cyc\biggr)_{\text{J}} \elnn
-&\frac{2}{7}b_{01}b_{10}^\La\biggl(2 \mathcal{S}_2(\k_1,\k_2)P(k_1)P(k_2)\left(\frac{1}{\alpha(k_1)}+\frac{1}{\alpha(k_2)}\right)+2\cyc\biggr)_{\text{K}} \elnn
-&b_{10}b_{01}^\La\biggl(4 P(k_1)P(k_2)\left(\frac{\mathcal{N}_2(\k_1,\k_2)}{\alpha(k_2)}+\frac{\mathcal{N}_2(\k_2,\k_1)}{\alpha(k_1)}\right)+2\cyc\biggr)_{\text{L}} \elnn
-&b_{01}b_{01}^\La\biggl(2 P(k_1)P(k_2)\left(\frac{\mathcal{N}_2(\k_1,\k_2)}{\alpha(k_2)}+\frac{\mathcal{N}_2(\k_2,\k_1)}{\alpha(k_1)}\right)\left(\frac{1}{\alpha(k_1)}+\frac{1}{\alpha(k_2)}\right)+2\cyc\biggr)_{\text{L}} \nonumber
\end{align}
The terms with label going from A to I match exactly the result of Eq.$(5.6)$ in BSS but, as for the halo bispectrum, new terms appear. J, K are due to the tidal term $s^2$: schematically, they generated by $\langle s^2 \delta^{(1)} \delta^{(1)}\rangle$, $\langle s^2 \delta^{(1)} \varphi \rangle$ respectively. We see by comparison with \cref{eq:bhhmg} that J is present regardless of PNG, while K comes from the coupling between the tidal term and $\varphi$ which is specifically introduced by PNG. L and M are present because of $n^2$ and, therefore, depend on the presence of PNG. These are schematically generated by $\langle n^2 \delta^{(1)} \delta^{(1)}\rangle$, $\langle n^2 \delta^{(1)} \varphi \rangle$,  respectively.

Finally, for the halo-matter-matter bispectrum when Gaussian initial conditions are assumed we find
\begin{align}
B_\text{hmm}(\k_1,\k_2,\k_3)=& b_{10}\biggl(6 \mathcal{F}_2(\k_1,\k_2)P(k_1)P(k_2)+2\cyc\biggr)_{\text{A}} \elnn
+&b_{20}\biggl(2P(k_1)P(k_2)+2\cyc\biggr)_{\text{C}} \\
-&\frac{2}{7}b_{10}^\La \biggl(2 \mathcal{S}_2(\k_1,\k_2)P(k_1)P(k_2)+2\cyc\biggr)_{\text{F}} \nonumber \,,
\end{align}
where the same considerations for \cref{eq:bhhm} apply. Then, assuming PNG, the halo-matter-matter bispectrum reads
\begin{align}
B_\text{hmm}(\k_1,\k_2,\k_3)=& b_{10}\biggl(6 \mathcal{F}_2(\k_1,\k_2)P(k_1)P(k_2)+6\fnl\alpha(k_3)\frac{P(k_1)P(k_2)}{\alpha(k_1)\alpha(k_2)}+2\cyc\biggr)_{\text{A}} \elnn
+&b_{01}\biggl(2 \mathcal{F}_2(\k_1,\k_2)P(k_1)P(k_2)\left(\frac{1}{\alpha(k_1)}+\frac{1}{\alpha(k_2)} \right) \elnn
+&2\fnl \alpha(k_3) P(k_1)P(k_2)\left(\frac{1}{\alpha^2(k_1)\alpha(k_2)} +\frac{1}{\alpha(k_1)\alpha^2(k_2)}\right)+2\cyc\biggr)_{\text{B}} \elnn
+&b_{20}\biggl(2P(k_1)P(k_2)+2\cyc\biggr)_{\text{C}} \label{eq:bhmm}\\
+&b_{11}\biggl(P(k_1)P(k_2)\left(\frac{1}{\alpha(k_1)}+\frac{1}{\alpha(k_2)}\right)+2\cyc\biggr)_{\text{D}} \elnn
+&b_{02}\biggl(2\frac{P(k_1)P(k_2)}{\alpha(k_1)\alpha(k_2)}+2\cyc\biggr)_{\text{E}} \elnn
-&\frac{2}{7}b_{10}^\La \biggl(2 \mathcal{S}_2(\k_1,\k_2)P(k_1)P(k_2)+2\cyc\biggr)_{\text{F}} \elnn
-&b_{01}^\La\biggl(2 P(k_1)P(k_2)\left(\frac{\mathcal{N}_2(\k_1,\k_2)}{\alpha(k_2)}+\frac{\mathcal{N}_2(\k_2,\k_1)}{\alpha(k_1)}\right)+2\cyc\biggr)_{\text{G}} \nonumber \,,
\end{align}
with the terms going from A to E matching exactly Eq.$(5.8)$ of BSS. As above, new terms appear: F is due to tidal term $s^2$ and it is generated by $\langle s^2 \delta^{(1)} \delta^{(1)}\rangle$, regardless of the presence of PNG, while G is due to $n^2$, sourced by $\langle n^2 \delta^{(1)} \delta^{(1)}\rangle$.

\bibliographystyle{JHEP}
\bibliography{biblio}

\providecommand{\href}[2]{#2}\begingroup\raggedright\begin{thebibliography}{100}

\bibitem{Bartolo:2004if}
N.~Bartolo, E.~Komatsu, S.~Matarrese, and A.~Riotto, {\it {Non-Gaussianity from
  inflation: Theory and observations}},  {\em Phys.Rept.} {\bf 402} (2004)
  103--266, [\href{http://arxiv.org/abs/astro-ph/0406398}{{\tt
  astro-ph/0406398}}].

\bibitem{Wands-2010review}
D.~{Wands}, {\it {Local non-Gaussianity from inflation}},  {\em Classical and
  Quantum Gravity} {\bf 27} (June, 2010) 124002,
  [\href{http://arxiv.org/abs/1004.0818}{{\tt arXiv:1004.0818}}].

\bibitem{Koyama:2010xj}
K.~Koyama, {\it {Non-Gaussianity of quantum fields during inflation}},  {\em
  Class.Quant.Grav.} {\bf 27} (2010) 124001,
  [\href{http://arxiv.org/abs/1002.0600}{{\tt arXiv:1002.0600}}].

\bibitem{Alvarez:2014vva}
M.~Alvarez, T.~Baldauf, J.~R. Bond, N.~Dalal, R.~de~Putter, et~al., {\it
  {Testing Inflation with Large Scale Structure: Connecting Hopes with
  Reality}},  \href{http://arxiv.org/abs/1412.4671}{{\tt arXiv:1412.4671}}.

\bibitem{WMAP9}
C.~L. {Bennett}, D.~{Larson}, J.~L. {Weiland}, N.~{Jarosik}, G.~{Hinshaw},
  N.~{Odegard}, K.~M. {Smith}, R.~S. {Hill}, B.~{Gold}, M.~{Halpern},
  E.~{Komatsu}, M.~R. {Nolta}, L.~{Page}, D.~N. {Spergel}, E.~{Wollack},
  J.~{Dunkley}, A.~{Kogut}, M.~{Limon}, S.~S. {Meyer}, G.~S. {Tucker}, and
  E.~L. {Wright}, {\it {Nine-year Wilkinson Microwave Anisotropy Probe (WMAP)
  Observations: Final Maps and Results}},  {\em \apjs} {\bf 208} (Oct., 2013)
  20, [\href{http://arxiv.org/abs/1212.5225}{{\tt arXiv:1212.5225}}].

\bibitem{Ade:2015ava}
{\bf Planck} Collaboration, P.~Ade et~al., {\it {Planck 2015 results. XVII.
  Constraints on primordial non-Gaussianity}},
  \href{http://arxiv.org/abs/1502.0159}{{\tt arXiv:1502.0159}}.

\bibitem{Andre:2013nfa}
{\bf PRISM} Collaboration, P.~Andr\'e et~al., {\it {PRISM (Polarized Radiation
  Imaging and Spectroscopy Mission): An Extended White Paper}},  {\em JCAP}
  {\bf 1402} (2014) 006, [\href{http://arxiv.org/abs/1310.1554}{{\tt
  arXiv:1310.1554}}].

\bibitem{Dalal:2007cu}
N.~Dalal, O.~Dore, D.~Huterer, and A.~Shirokov, {\it {The imprints of
  primordial non-gaussianities on large-scale structure: scale dependent bias
  and abundance of virialized objects}},  {\em Phys.Rev.} {\bf D77} (2008)
  123514, [\href{http://arxiv.org/abs/0710.4560}{{\tt arXiv:0710.4560}}].

\bibitem{Matarrese:2008nc}
S.~Matarrese and L.~Verde, {\it {The effect of primordial non-Gaussianity on
  halo bias}},  {\em Astrophys.J.} {\bf 677} (2008) L77--L80,
  [\href{http://arxiv.org/abs/0801.4826}{{\tt arXiv:0801.4826}}].

\bibitem{Slosar:2008hx}
A.~Slosar, C.~Hirata, U.~Seljak, S.~Ho, and N.~Padmanabhan, {\it {Constraints
  on local primordial non-Gaussianity from large scale structure}},  {\em JCAP}
  {\bf 0808} (2008) 031, [\href{http://arxiv.org/abs/0805.3580}{{\tt
  arXiv:0805.3580}}].

\bibitem{Afshordi:2008ru}
N.~Afshordi and A.~J. Tolley, {\it {Primordial non-gaussianity, statistics of
  collapsed objects, and the Integrated Sachs-Wolfe effect}},  {\em Phys.Rev.}
  {\bf D78} (2008) 123507, [\href{http://arxiv.org/abs/0806.1046}{{\tt
  arXiv:0806.1046}}].

\bibitem{Desjacques:2008vf}
V.~Desjacques, U.~Seljak, and I.~Iliev, {\it {Scale-dependent bias induced by
  local non-Gaussianity: A comparison to N-body simulations}},  {\em
  Mon.Not.Roy.Astron.Soc.} {\bf 396} (2009) 85--96,
  [\href{http://arxiv.org/abs/0811.2748}{{\tt arXiv:0811.2748}}].

\bibitem{Desjacques1}
V.~{Desjacques}, D.~{Jeong}, and F.~{Schmidt}, {\it {Non-Gaussian Halo Bias
  Re-examined: Mass-dependent Amplitude from the Peak-Background Split and
  Thresholding}},  {\em \prd} {\bf 84} (Sept., 2011) 063512,
  [\href{http://arxiv.org/abs/1105.3628}{{\tt arXiv:1105.3628}}].

\bibitem{Desjacques2}
V.~{Desjacques}, D.~{Jeong}, and F.~{Schmidt}, {\it {Accurate predictions for
  the scale-dependent galaxy bias from primordial non-Gaussianity}},  {\em
  \prd} {\bf 84} (Sept., 2011) 061301,
  [\href{http://arxiv.org/abs/1105.3476}{{\tt arXiv:1105.3476}}].

\bibitem{Gong}
J.-O. {Gong} and S.~{Yokoyama}, {\it {Scale-dependent bias from primordial
  non-Gaussianity with trispectrum}},  {\em \mnras} {\bf 417} (Oct., 2011)
  L79--L82, [\href{http://arxiv.org/abs/1106.4404}{{\tt arXiv:1106.4404}}].

\bibitem{Huterer2013}
D.~{Huterer}, C.~E. {Cunha}, and W.~{Fang}, {\it {Calibration errors unleashed:
  effects on cosmological parameters and requirements for large-scale structure
  surveys}},  {\em \mnras} {\bf 432} (July, 2013) 2945--2961,
  [\href{http://arxiv.org/abs/1211.1015}{{\tt arXiv:1211.1015}}].

\bibitem{Pullen2013}
A.~R. {Pullen} and C.~M. {Hirata}, {\it {Systematic Effects in Large-Scale
  Angular Power Spectra of Photometric Quasars and Implications for
  Constraining Primordial Non-Gaussianity}},  {\em \pasp} {\bf 125} (June,
  2013) 705--718, [\href{http://arxiv.org/abs/1212.4500}{{\tt
  arXiv:1212.4500}}].

\bibitem{Agarwal:2013ajb}
N.~Agarwal, S.~Ho, A.~D. Myers, H.-J. Seo, A.~J. Ross, et~al., {\it
  {Characterizing unknown systematics in large scale structure surveys}},  {\em
  JCAP} {\bf 1404} (2014) 007, [\href{http://arxiv.org/abs/1309.2954}{{\tt
  arXiv:1309.2954}}].

\bibitem{Leistedt:2014zqa}
B.~Leistedt, H.~V. Peiris, and N.~Roth, {\it {Constraints on Primordial
  Non-Gaussianity from 800 000 Photometric Quasars}},  {\em Phys.Rev.Lett.}
  {\bf 113} (2014), no.~22 221301, [\href{http://arxiv.org/abs/1405.4315}{{\tt
  arXiv:1405.4315}}].

\bibitem{Ross:2013}
A.~J. {Ross}, W.~J. {Percival}, A.~{Carnero}, G.-b. {Zhao}, M.~{Manera},
  A.~{Raccanelli}, E.~{Aubourg}, D.~{Bizyaev}, H.~{Brewington}, J.~{Brinkmann},
  J.~R. {Brownstein}, A.~J. {Cuesta}, L.~A.~N. {da Costa}, D.~J. {Eisenstein},
  G.~{Ebelke}, H.~{Guo}, J.-C. {Hamilton}, M.~V. {Maga{\~n}a},
  E.~{Malanushenko}, V.~{Malanushenko}, C.~{Maraston}, F.~{Montesano}, R.~C.
  {Nichol}, D.~{Oravetz}, K.~{Pan}, F.~{Prada}, A.~G. {S{\'a}nchez},
  L.~{Samushia}, D.~J. {Schlegel}, D.~P. {Schneider}, H.-J. {Seo},
  A.~{Sheldon}, A.~{Simmons}, S.~{Snedden}, M.~E.~C. {Swanson}, D.~{Thomas},
  J.~L. {Tinker}, R.~{Tojeiro}, and I.~{Zehavi}, {\it {The clustering of
  galaxies in the SDSS-III DR9 Baryon Oscillation Spectroscopic Survey:
  constraints on primordial non-Gaussianity}},  {\em \mnras} {\bf 428} (Jan.,
  2013) 1116--1127, [\href{http://arxiv.org/abs/1208.1491}{{\tt
  arXiv:1208.1491}}].

\bibitem{Karagiannis:2013xea}
D.~Karagiannis, T.~Shanks, and N.~P. Ross, {\it {Search for primordial
  non-Gaussianity in the quasars of SDSS-III BOSS DR9}},  {\em
  Mon.Not.Roy.Astron.Soc.} {\bf 441} (2014) 486--502,
  [\href{http://arxiv.org/abs/1310.6716}{{\tt arXiv:1310.6716}}].

\bibitem{Seljak:2008xr}
U.~Seljak, {\it {Extracting primordial non-gaussianity without cosmic
  variance}},  {\em Phys.Rev.Lett.} {\bf 102} (2009) 021302,
  [\href{http://arxiv.org/abs/0807.1770}{{\tt arXiv:0807.1770}}].

\bibitem{GilMarin10}
H.~{Gil-Mar{\'{\i}}n}, C.~{Wagner}, L.~{Verde}, R.~{Jimenez}, and A.~F.
  {Heavens}, {\it {Reducing sample variance: halo biasing, non-linearity and
  stochasticity}},  {\em \mnras} {\bf 407} (Sept., 2010) 772--790,
  [\href{http://arxiv.org/abs/1003.3238}{{\tt arXiv:1003.3238}}].

\bibitem{Hamaus11}
N.~{Hamaus}, U.~{Seljak}, and V.~{Desjacques}, {\it {Optimal constraints on
  local primordial non-Gaussianity from the two-point statistics of large-scale
  structure}},  {\em \prd} {\bf 84} (Oct., 2011) 083509,
  [\href{http://arxiv.org/abs/1104.2321}{{\tt arXiv:1104.2321}}].

\bibitem{Biagetti13}
M.~{Biagetti}, V.~{Desjacques}, and A.~{Riotto}, {\it {Testing multifield
  inflation with halo bias}},  {\em \mnras} {\bf 429} (Feb., 2013) 1774--1780,
  [\href{http://arxiv.org/abs/1208.1616}{{\tt arXiv:1208.1616}}].

\bibitem{Ferramacho:2014pua}
L.~D. Ferramacho, M.~G. Santos, M.~J. Jarvis, and S.~Camera, {\it {Radio Galaxy
  populations and the multi-tracer technique: pushing the limits on primordial
  non-Gaussianity}},  {\em Mon.Not.Roy.Astron.Soc.} {\bf 442} (2014) 2511,
  [\href{http://arxiv.org/abs/1402.2290}{{\tt arXiv:1402.2290}}].

\bibitem{Yamauchi:2014ioa}
D.~Yamauchi, K.~Takahashi, and M.~Oguri, {\it {Constraining primordial
  non-Gaussianity via a multitracer technique with surveys by Euclid and Square
  Kilometre Array}},  {\em Phys.Rev.} {\bf D90} (2014) 083520,
  [\href{http://arxiv.org/abs/1407.5453}{{\tt arXiv:1407.5453}}].

\bibitem{Seljak09}
U.~{Seljak}, N.~{Hamaus}, and V.~{Desjacques}, {\it {How to Suppress the Shot
  Noise in Galaxy Surveys}},  {\em Physical Review Letters} {\bf 103} (Aug.,
  2009) 091303, [\href{http://arxiv.org/abs/0904.2963}{{\tt arXiv:0904.2963}}].

\bibitem{Hamaus10}
N.~{Hamaus}, U.~{Seljak}, V.~{Desjacques}, R.~E. {Smith}, and T.~{Baldauf},
  {\it {Minimizing the stochasticity of halos in large-scale structure
  surveys}},  {\em \prd} {\bf 82} (Aug., 2010) 043515,
  [\href{http://arxiv.org/abs/1004.5377}{{\tt arXiv:1004.5377}}].

\bibitem{Xia1:2010}
J.-Q. {Xia}, M.~{Viel}, C.~{Baccigalupi}, G.~{De Zotti}, S.~{Matarrese}, and
  L.~{Verde}, {\it {Primordial Non-Gaussianity and the NRAO VLA Sky Survey}},
  {\em \apjl} {\bf 717} (July, 2010) L17--L21,
  [\href{http://arxiv.org/abs/1003.3451}{{\tt arXiv:1003.3451}}].

\bibitem{Xia2:2010}
J.-Q. {Xia}, A.~{Bonaldi}, C.~{Baccigalupi}, G.~{De Zotti}, S.~{Matarrese},
  L.~{Verde}, and M.~{Viel}, {\it {Constraining primordial non-Gaussianity with
  high-redshift probes}},  {\em \jcap} {\bf 8} (Aug., 2010) 13,
  [\href{http://arxiv.org/abs/1007.1969}{{\tt arXiv:1007.1969}}].

\bibitem{Giannantonio:2013uqa}
T.~Giannantonio, A.~J. Ross, W.~J. Percival, R.~Crittenden, D.~Bacher, et~al.,
  {\it {Improved Primordial Non-Gaussianity Constraints from Measurements of
  Galaxy Clustering and the Integrated Sachs-Wolfe Effect}},  {\em Phys.Rev.}
  {\bf D89} (2014) 023511, [\href{http://arxiv.org/abs/1303.1349}{{\tt
  arXiv:1303.1349}}].

\bibitem{Ho:2013lda}
S.~Ho, N.~Agarwal, A.~D. Myers, R.~Lyons, A.~Disbrow, et~al., {\it {Sloan
  Digital Sky Survey III Photometric Quasar Clustering: Probing the Initial
  Conditions of the Universe using the Largest Volume}},
  \href{http://arxiv.org/abs/1311.2597}{{\tt arXiv:1311.2597}}.

\bibitem{Giannantonio:2013kqa}
T.~Giannantonio and W.~J. Percival, {\it {Using correlations between CMB
  lensing and large-scale structure to measure primordial non-Gaussianity}},
  {\em Mon.Not.Roy.Astron.Soc.} {\bf 441} (2014) L16–L20,
  [\href{http://arxiv.org/abs/1312.5154}{{\tt arXiv:1312.5154}}].

\bibitem{Giannantonio:2011}
T.~{Giannantonio}, C.~{Porciani}, J.~{Carron}, A.~{Amara}, and A.~{Pillepich},
  {\it {Constraining primordial non-Gaussianity with future galaxy surveys}},
  {\em \mnras} {\bf 422} (June, 2012) 2854--2877,
  [\href{http://arxiv.org/abs/1109.0958}{{\tt arXiv:1109.0958}}].

\bibitem{Raccanelli:2014kga}
A.~Raccanelli, O.~Doré, D.~J. Bacon, R.~Maartens, M.~G. Santos, et~al., {\it
  {Probing primordial non-Gaussianity via iSW measurements with SKA continuum
  surveys}},  \href{http://arxiv.org/abs/1406.0010}{{\tt arXiv:1406.0010}}.

\bibitem{Camera:2014bwa}
S.~Camera, M.~G. Santos, and R.~Maartens, {\it {Probing primordial
  non-Gaussianity with SKA galaxy redshift surveys: a fully relativistic
  analysis}},  \href{http://arxiv.org/abs/1409.8286}{{\tt arXiv:1409.8286}}.

\bibitem{dePutter:2014lna}
R.~de~Putter and O.~Doré, {\it {Designing an Inflation Galaxy Survey: how to
  measure $\sigma(f_{\rm NL}) \sim 1$ using scale-dependent galaxy bias}},
  \href{http://arxiv.org/abs/1412.3854}{{\tt arXiv:1412.3854}}.

\bibitem{Leistedt:2014wia}
B.~Leistedt and H.~V. Peiris, {\it {Exploiting the full potential of
  photometric quasar surveys: Optimal power spectra through blind mitigation of
  systematics}},  {\em Mon.Not.Roy.Astron.Soc.} {\bf 444} (2014) 2,
  [\href{http://arxiv.org/abs/1404.6530}{{\tt arXiv:1404.6530}}].

\bibitem{Roth:2012}
N.~{Roth} and C.~{Porciani}, {\it {Can we really measure f$_{NL}$ from the
  galaxy power spectrum?}},  {\em \mnras} {\bf 425} (Sept., 2012) L81--L85,
  [\href{http://arxiv.org/abs/1205.3165}{{\tt arXiv:1205.3165}}].

\bibitem{Mao:2014caa}
Q.~Mao, A.~A. Berlind, C.~K. McBride, R.~J. Scherrer, R.~Scoccimarro, et~al.,
  {\it {Constraining Primordial Non-Gaussianity with Moments of the Large Scale
  Density Field}},  {\em Mon.Not.Roy.Astron.Soc.} {\bf 443} (2014) 1402--1415,
  [\href{http://arxiv.org/abs/1404.3725}{{\tt arXiv:1404.3725}}].

\bibitem{Scoccimarro:2003wn}
R.~Scoccimarro, E.~Sefusatti, and M.~Zaldarriaga, {\it {Probing primordial
  non-Gaussianity with large - scale structure}},  {\em Phys.Rev.} {\bf D69}
  (2004) 103513, [\href{http://arxiv.org/abs/astro-ph/0312286}{{\tt
  astro-ph/0312286}}].

\bibitem{Sefusatti:2007ih}
E.~Sefusatti and E.~Komatsu, {\it {The bispectrum of galaxies from
  high-redshift galaxy surveys: Primordial non-Gaussianity and non-linear
  galaxy bias}},  {\em Phys.Rev.} {\bf D76} (2007) 083004,
  [\href{http://arxiv.org/abs/0705.0343}{{\tt arXiv:0705.0343}}].

\bibitem{Baldauf:2010vn}
T.~Baldauf, U.~Seljak, and L.~Senatore, {\it {Primordial non-Gaussianity in the
  Bispectrum of the Halo Density Field}},  {\em JCAP} {\bf 1104} (2011) 006,
  [\href{http://arxiv.org/abs/1011.1513}{{\tt arXiv:1011.1513}}].

\bibitem{Jeong:2009}
D.~{Jeong} and E.~{Komatsu}, {\it {Primordial Non-Gaussianity, Scale-dependent
  Bias, and the Bispectrum of Galaxies}},  {\em \apj} {\bf 703} (Oct., 2009)
  1230--1248, [\href{http://arxiv.org/abs/0904.0497}{{\tt arXiv:0904.0497}}].

\bibitem{Tasinato:2013vna}
G.~Tasinato, M.~Tellarini, A.~J. Ross, and D.~Wands, {\it {Primordial
  non-Gaussianity in the bispectra of large-scale structure}},  {\em JCAP} {\bf
  1403} (2014) 032, [\href{http://arxiv.org/abs/1310.7482}{{\tt
  arXiv:1310.7482}}].

\bibitem{Scoccimarro:2000sp}
R.~Scoccimarro, H.~A. Feldman, J.~N. Fry, and J.~A. Frieman, {\it {The
  Bispectrum of IRAS redshift catalogs}},  {\em Astrophys.J.} {\bf 546} (2001)
  652, [\href{http://arxiv.org/abs/astro-ph/0004087}{{\tt astro-ph/0004087}}].

\bibitem{Verde:2001sf}
L.~Verde, A.~F. Heavens, W.~J. Percival, S.~Matarrese, C.~M. Baugh, et~al.,
  {\it {The 2dF Galaxy Redshift Survey: The Bias of galaxies and the density of
  the Universe}},  {\em Mon.Not.Roy.Astron.Soc.} {\bf 335} (2002) 432,
  [\href{http://arxiv.org/abs/astro-ph/0112161}{{\tt astro-ph/0112161}}].

\bibitem{Jing:2003nb}
Y.~Jing and G.~Boerner, {\it {The three-point correlation function of galaxies
  determined from the 2df galaxy redshift survey}},  {\em Astrophys.J.} {\bf
  607} (2004) 140--163, [\href{http://arxiv.org/abs/astro-ph/0311585}{{\tt
  astro-ph/0311585}}].

\bibitem{Gaztanaga:2005an}
E.~Gaztanaga, P.~Norberg, C.~Baugh, and D.~Croton, {\it {Statistical analysis
  of galaxy surveys. 2. The 3-point galaxy correlation function measured from
  the 2dFGRS}},  {\em Mon.Not.Roy.Astron.Soc.} {\bf 364} (2005) 620--634,
  [\href{http://arxiv.org/abs/astro-ph/0506249}{{\tt astro-ph/0506249}}].

\bibitem{FMarin:2011}
F.~{Mar{\'{\i}}n}, {\it {The Large-scale Three-point Correlation Function of
  Sloan Digital Sky Survey Luminous Red Galaxies}},  {\em \apj} {\bf 737}
  (Aug., 2011) 97, [\href{http://arxiv.org/abs/1011.4530}{{\tt
  arXiv:1011.4530}}].

\bibitem{McBride:2011}
C.~K. {McBride}, A.~J. {Connolly}, J.~P. {Gardner}, R.~{Scranton},
  R.~{Scoccimarro}, A.~A. {Berlind}, F.~{Mar{\'{\i}}n}, and D.~P. {Schneider},
  {\it {Three-point Correlation Functions of SDSS Galaxies: Constraining
  Galaxy-mass Bias}},  {\em \apj} {\bf 739} (Oct., 2011) 85,
  [\href{http://arxiv.org/abs/1012.3462}{{\tt arXiv:1012.3462}}].

\bibitem{Marin:2013bbb}
{\bf WiggleZ} Collaboration, F.~A. Marin et~al., {\it {The WiggleZ Dark Energy
  Survey: constraining galaxy bias and cosmic growth with 3-point correlation
  functions}},  {\em Mon.Not.Roy.Astron.Soc.} {\bf 432} (2013) 2654,
  [\href{http://arxiv.org/abs/1303.6644}{{\tt arXiv:1303.6644}}].

\bibitem{Gil-Marin:2014sta}
H.~{Gil-Mar{\'{\i}}n}, J.~Noreña, L.~Verde, W.~J. Percival, C.~Wagner, et~al.,
  {\it {The power spectrum and bispectrum of SDSS DR11 BOSS galaxies I: bias
  and gravity}},  \href{http://arxiv.org/abs/1407.5668}{{\tt arXiv:1407.5668}}.

\bibitem{Gil-Marin:2014baa}
H.~{Gil-Mar{\'{\i}}n}, L.~Verde, J.~Noreña, A.~J. Cuesta, L.~Samushia, et~al.,
  {\it {The power spectrum and bispectrum of SDSS DR11 BOSS galaxies II:
  cosmological interpretation}},  \href{http://arxiv.org/abs/1408.0027}{{\tt
  arXiv:1408.0027}}.

\bibitem{Fry:1992vr}
J.~N. Fry and E.~Gaztanaga, {\it {Biasing and hierarchical statistics in large
  scale structure}},  {\em Astrophys.J.} {\bf 413} (1993) 447--452,
  [\href{http://arxiv.org/abs/astro-ph/9302009}{{\tt astro-ph/9302009}}].

\bibitem{Baldauf:2012hs}
T.~Baldauf, U.~Seljak, V.~Desjacques, and P.~McDonald, {\it {Evidence for
  Quadratic Tidal Tensor Bias from the Halo Bispectrum}},  {\em Phys.Rev.} {\bf
  D86} (2012) 083540, [\href{http://arxiv.org/abs/1201.4827}{{\tt
  arXiv:1201.4827}}].

\bibitem{Saito:2014qha}
S.~Saito, T.~Baldauf, Z.~Vlah, U.~Seljak, T.~Okumura, et~al., {\it
  {Understanding higher-order nonlocal halo bias at large scales by combining
  the power spectrum with the bispectrum}},  {\em Phys.Rev.} {\bf D90} (2014),
  no.~12 123522, [\href{http://arxiv.org/abs/1405.1447}{{\tt
  arXiv:1405.1447}}].

\bibitem{Chan:2012}
K.~C. {Chan}, R.~{Scoccimarro}, and R.~K. {Sheth}, {\it {Gravity and
  large-scale nonlocal bias}},  {\em \prd} {\bf 85} (Apr., 2012) 083509,
  [\href{http://arxiv.org/abs/1201.3614}{{\tt arXiv:1201.3614}}].

\bibitem{Roth:2011}
N.~{Roth} and C.~{Porciani}, {\it {Testing standard perturbation theory and the
  Eulerian local biasing scheme against N-body simulations}},  {\em \mnras}
  {\bf 415} (July, 2011) 829--844, [\href{http://arxiv.org/abs/1101.1520}{{\tt
  arXiv:1101.1520}}].

\bibitem{Pollack:2013alj}
J.~E. Pollack, R.~E. Smith, and C.~Porciani, {\it {A new method to measure
  galaxy bias}},  \href{http://arxiv.org/abs/1309.0504}{{\tt arXiv:1309.0504}}.

\bibitem{2009JCAP...08..020M}
P.~{McDonald} and A.~{Roy}, {\it {Clustering of dark matter tracers:
  generalizing bias for the coming era of precision LSS}},  {\em \jcap} {\bf 8}
  (Aug., 2009) 20, [\href{http://arxiv.org/abs/0902.0991}{{\tt
  arXiv:0902.0991}}].

\bibitem{Mo:1995cs}
H.~Mo and S.~D. White, {\it {An Analytic model for the spatial clustering of
  dark matter halos}},  {\em Mon.Not.Roy.Astron.Soc.} {\bf 282} (1996) 347,
  [\href{http://arxiv.org/abs/astro-ph/9512127}{{\tt astro-ph/9512127}}].

\bibitem{Sheth:2013}
R.~K. {Sheth}, K.~C. {Chan}, and R.~{Scoccimarro}, {\it {Nonlocal Lagrangian
  bias}},  {\em \prd} {\bf 87} (Apr., 2013) 083002,
  [\href{http://arxiv.org/abs/1207.7117}{{\tt arXiv:1207.7117}}].

\bibitem{Matsubara:2011PhRvD}
T.~{Matsubara}, {\it {Nonlinear perturbation theory integrated with nonlocal
  bias, redshift-space distortions, and primordial non-Gaussianity}},  {\em
  \prd} {\bf 83} (Apr., 2011) 083518,
  [\href{http://arxiv.org/abs/1102.4619}{{\tt arXiv:1102.4619}}].

\bibitem{Catelan:2000vn}
P.~Catelan, C.~Porciani, and M.~Kamionkowski, {\it {Two ways of biasing galaxy
  formation}},  {\em Mon.Not.Roy.Astron.Soc.} {\bf 318} (2000) 39,
  [\href{http://arxiv.org/abs/astro-ph/0005544}{{\tt astro-ph/0005544}}].

\bibitem{Giannantonio:2009ak}
T.~Giannantonio and C.~Porciani, {\it {Structure formation from non-Gaussian
  initial conditions: multivariate biasing, statistics, and comparison with
  N-body simulations}},  {\em Phys.Rev.} {\bf D81} (2010) 063530,
  [\href{http://arxiv.org/abs/0911.0017}{{\tt arXiv:0911.0017}}].

\bibitem{Sefusatti:2011}
E.~{Sefusatti}, M.~{Crocce}, and V.~{Desjacques}, {\it {The matter bispectrum
  in N-body simulations with non-Gaussian initial conditions}},  {\em \mnras}
  {\bf 406} (Aug., 2010) 1014--1028,
  [\href{http://arxiv.org/abs/1003.0007}{{\tt arXiv:1003.0007}}].

\bibitem{Scoccimarro:2012}
R.~{Scoccimarro}, L.~{Hui}, M.~{Manera}, and K.~C. {Chan}, {\it {Large-scale
  bias and efficient generation of initial conditions for nonlocal primordial
  non-Gaussianity}},  {\em \prd} {\bf 85} (Apr., 2012) 083002,
  [\href{http://arxiv.org/abs/1108.5512}{{\tt arXiv:1108.5512}}].

\bibitem{Yokoyama:2013mta}
S.~Yokoyama, T.~Matsubara, and A.~Taruya, {\it {Halo/galaxy bispectrum with
  primordial non-Gaussianity from integrated perturbation theory}},  {\em
  Phys.Rev.} {\bf D89} (2014), no.~4 043524,
  [\href{http://arxiv.org/abs/1310.4925}{{\tt arXiv:1310.4925}}].

\bibitem{Gangui:1993tt}
A.~Gangui, F.~Lucchin, S.~Matarrese, and S.~Mollerach, {\it {The Three point
  correlation function of the cosmic microwave background in inflationary
  models}},  {\em Astrophys.J.} {\bf 430} (1994) 447--457,
  [\href{http://arxiv.org/abs/astro-ph/9312033}{{\tt astro-ph/9312033}}].

\bibitem{Verde:1999ij}
L.~Verde, L.-M. Wang, A.~Heavens, and M.~Kamionkowski, {\it {Large scale
  structure, the cosmic microwave background, and primordial non-gaussianity}},
   {\em Mon.Not.Roy.Astron.Soc.} {\bf 313} (2000) L141--L147,
  [\href{http://arxiv.org/abs/astro-ph/9906301}{{\tt astro-ph/9906301}}].

\bibitem{Komatsu:2001rj}
E.~Komatsu and D.~N. Spergel, {\it {Acoustic signatures in the primary
  microwave background bispectrum}},  {\em Phys.Rev.} {\bf D63} (2001) 063002,
  [\href{http://arxiv.org/abs/astro-ph/0005036}{{\tt astro-ph/0005036}}].

\bibitem{Bartolo2010}
N.~{Bartolo}, S.~{Matarrese}, O.~{Pantano}, and A.~{Riotto}, {\it {Second-order
  matter perturbations in a {$\Lambda$}CDM cosmology and non-Gaussianity}},
  {\em Classical and Quantum Gravity} {\bf 27} (June, 2010) 124009,
  [\href{http://arxiv.org/abs/1002.3759}{{\tt arXiv:1002.3759}}].

\bibitem{Bruni:2013qta}
M.~Bruni, J.~C. Hidalgo, N.~Meures, and D.~Wands, {\it {Non-Gaussian Initial
  Conditions in ΛCDM: Newtonian, Relativistic, and Primordial Contributions}},
   {\em Astrophys.J.} {\bf 785} (2014) 2,
  [\href{http://arxiv.org/abs/1307.1478}{{\tt arXiv:1307.1478}}].

\bibitem{Bruni:2014xma}
M.~Bruni, J.~C. Hidalgo, and D.~Wands, {\it {Einstein's signature in
  cosmological large-scale structure}},  {\em Astrophys.J.} {\bf 794} (2014),
  no.~1 L11, [\href{http://arxiv.org/abs/1405.7006}{{\tt arXiv:1405.7006}}].

\bibitem{camb}
\url{http://camb.info/}.

\bibitem{Ade:2013ydc}
P.~Ade et~al., {\it {Planck 2013 Results. XXIV. Constraints on primordial
  non-Gaussianity}},  {\em Astron.Astrophys.} {\bf 571} (2014) A24,
  [\href{http://arxiv.org/abs/1303.5084}{{\tt arXiv:1303.5084}}].

\bibitem{Bernardeau:2001qr}
F.~Bernardeau, S.~Colombi, E.~Gaztanaga, and R.~Scoccimarro, {\it {Large scale
  structure of the universe and cosmological perturbation theory}},  {\em
  Phys.Rept.} {\bf 367} (2002) 1--248,
  [\href{http://arxiv.org/abs/astro-ph/0112551}{{\tt astro-ph/0112551}}].

\bibitem{Bertacca:2015mca}
D.~Bertacca, N.~Bartolo, M.~Bruni, K.~Koyama, R.~Maartens, et~al., {\it {Galaxy
  bias and gauges at second order in General Relativity}},
  \href{http://arxiv.org/abs/1501.0316}{{\tt arXiv:1501.0316}}.

\bibitem{Schmittfull:2014tca}
M.~Schmittfull, T.~Baldauf, and U.~Seljak, {\it {Near optimal bispectrum
  estimators for large-scale structure}},  {\em Phys.Rev.} {\bf D91} (2015),
  no.~4 043530, [\href{http://arxiv.org/abs/1411.6595}{{\tt arXiv:1411.6595}}].

\bibitem{peebles1980large}
P.~Peebles, {\em The Large-scale Structure of the Universe}.
\newblock Princeton series in physics. Princeton University Press, 1980.

\bibitem{Kaiser:1984sw}
N.~Kaiser, {\it {On the Spatial correlations of Abell clusters}},  {\em
  Astrophys.J.} {\bf 284} (1984) L9--L12.

\bibitem{Baumann:2013}
D.~{Baumann}, S.~{Ferraro}, D.~{Green}, and K.~M. {Smith}, {\it {Stochastic
  bias from non-Gaussian initial conditions}},  {\em \jcap} {\bf 5} (May, 2013)
  1, [\href{http://arxiv.org/abs/1209.2173}{{\tt arXiv:1209.2173}}].

\bibitem{Scoccimarro:2000gm}
R.~Scoccimarro, R.~K. Sheth, L.~Hui, and B.~Jain, {\it {How many galaxies fit
  in a halo? Constraints on galaxy formation efficiency from spatial
  clustering}},  {\em Astrophys.J.} {\bf 546} (2001) 20--34,
  [\href{http://arxiv.org/abs/astro-ph/0006319}{{\tt astro-ph/0006319}}].

\bibitem{Catelan:1997qw}
P.~Catelan, F.~Lucchin, S.~Matarrese, and C.~Porciani, {\it {The bias field of
  dark matter halos}},  {\em Mon.Not.Roy.Astron.Soc.} {\bf 297} (1998)
  692--712, [\href{http://arxiv.org/abs/astro-ph/9708067}{{\tt
  astro-ph/9708067}}].

\bibitem{Scoccimarro:2000ee}
R.~Scoccimarro and H.~Couchman, {\it {A fitting formula for the nonlinear
  evolution of the bispectrum}},  {\em Mon.Not.Roy.Astron.Soc.} {\bf 325}
  (2001) 1312, [\href{http://arxiv.org/abs/astro-ph/0009427}{{\tt
  astro-ph/0009427}}].

\bibitem{Marin:2012}
H.~{Gil-Mar{\'{\i}}n}, C.~{Wagner}, F.~{Fragkoudi}, R.~{Jimenez}, and
  L.~{Verde}, {\it {An improved fitting formula for the dark matter
  bispectrum}},  {\em \jcap} {\bf 2} (Feb., 2012) 47,
  [\href{http://arxiv.org/abs/1111.4477}{{\tt arXiv:1111.4477}}].

\bibitem{Sefusatti:2012}
E.~{Sefusatti}, M.~{Crocce}, and V.~{Desjacques}, {\it {The halo bispectrum in
  N-body simulations with non-Gaussian initial conditions}},  {\em \mnras} {\bf
  425} (Oct., 2012) 2903--2930, [\href{http://arxiv.org/abs/1111.6966}{{\tt
  arXiv:1111.6966}}].

\bibitem{EUCLID}
R.~{Laureijs}, J.~{Amiaux}, S.~{Arduini}, J.~. {Augu{\`e}res}, J.~{Brinchmann},
  R.~{Cole}, M.~{Cropper}, C.~{Dabin}, L.~{Duvet}, A.~{Ealet}, and et~al., {\it
  {Euclid Definition Study Report}},  {\em ArXiv e-prints} (Oct., 2011)
  [\href{http://arxiv.org/abs/1110.3193}{{\tt arXiv:1110.3193}}].

\bibitem{1974ApJ...187..425P}
W.~H. {Press} and P.~{Schechter}, {\it {Formation of Galaxies and Clusters of
  Galaxies by Self-Similar Gravitational Condensation}},  {\em \apj} {\bf 187}
  (Feb., 1974) 425--438.

\bibitem{Bardeen:1986}
J.~M. {Bardeen}, J.~R. {Bond}, N.~{Kaiser}, and A.~S. {Szalay}, {\it {The
  statistics of peaks of Gaussian random fields}},  {\em \apj} {\bf 304} (May,
  1986) 15--61.

\bibitem{Bond:1991}
J.~R. {Bond}, S.~{Cole}, G.~{Efstathiou}, and N.~{Kaiser}, {\it {Excursion set
  mass functions for hierarchical Gaussian fluctuations}},  {\em \apj} {\bf
  379} (Oct., 1991) 440--460.

\bibitem{Zentner:2006vw}
A.~R. Zentner, {\it {The Excursion Set Theory of Halo Mass Functions, Halo
  Clustering, and Halo Growth}},  {\em Int.J.Mod.Phys.} {\bf D16} (2007)
  763--816, [\href{http://arxiv.org/abs/astro-ph/0611454}{{\tt
  astro-ph/0611454}}].

\bibitem{Kitayama:1996ne}
T.~Kitayama and Y.~Suto, {\it {Semianalytical predictions for statistical
  properties of x-ray clusters of galaxies in cold dark matter universes}},
  {\em Astrophys.J.} {\bf 469} (1996) 480,
  [\href{http://arxiv.org/abs/astro-ph/9604141}{{\tt astro-ph/9604141}}].

\bibitem{Eke:1996ds}
V.~R. Eke, S.~Cole, and C.~S. Frenk, {\it {Using the evolution of clusters to
  constrain Omega}},  {\em Mon.Not.Roy.Astron.Soc.} {\bf 282} (1996) 263--280,
  [\href{http://arxiv.org/abs/astro-ph/9601088}{{\tt astro-ph/9601088}}].

\bibitem{Sheth:1999}
R.~K. {Sheth} and G.~{Tormen}, {\it {Large-scale bias and the peak background
  split}},  {\em \mnras} {\bf 308} (Sept., 1999) 119--126,
  [\href{http://arxiv.org/abs/astro-ph/9901122}{{\tt astro-ph/9901122}}].

\bibitem{Sheth:2001}
R.~K. {Sheth}, H.~J. {Mo}, and G.~{Tormen}, {\it {Ellipsoidal collapse and an
  improved model for the number and spatial distribution of dark matter
  haloes}},  {\em \mnras} {\bf 323} (May, 2001) 1--12,
  [\href{http://arxiv.org/abs/astro-ph/9907024}{{\tt astro-ph/9907024}}].

\bibitem{Sheth:2002}
R.~K. {Sheth} and G.~{Tormen}, {\it {An excursion set model of hierarchical
  clustering: ellipsoidal collapse and the moving barrier}},  {\em \mnras} {\bf
  329} (Jan., 2002) 61--75, [\href{http://arxiv.org/abs/astro-ph/0105113}{{\tt
  astro-ph/0105113}}].

\bibitem{LoVerde:2007ri}
M.~LoVerde, A.~Miller, S.~Shandera, and L.~Verde, {\it {Effects of
  Scale-Dependent Non-Gaussianity on Cosmological Structures}},  {\em JCAP}
  {\bf 0804} (2008) 014, [\href{http://arxiv.org/abs/0711.4126}{{\tt
  arXiv:0711.4126}}].

\bibitem{2011JCAP...08..003L}
M.~{LoVerde} and K.~M. {Smith}, {\it {The non-Gaussian halo mass function with
  f$_{NL}$, g$_{NL}$ and {$\tau$}$_{NL}$}},  {\em \jcap} {\bf 8} (Aug., 2011)
  3, [\href{http://arxiv.org/abs/1102.1439}{{\tt arXiv:1102.1439}}].

\bibitem{Suyama:2007bg}
T.~Suyama and M.~Yamaguchi, {\it {Non-Gaussianity in the modulated reheating
  scenario}},  {\em Phys.Rev.} {\bf D77} (2008) 023505,
  [\href{http://arxiv.org/abs/0709.2545}{{\tt arXiv:0709.2545}}].

\bibitem{Smith:2011ub}
K.~M. {Smith}, S.~{Ferraro}, and M.~{LoVerde}, {\it {Halo clustering and
  g$_{NL}$-type primordial non-gaussianity}},  {\em JCAP} {\bf 1203} (2012)
  032, [\href{http://arxiv.org/abs/1106.0503}{{\tt arXiv:1106.0503}}].

\end{thebibliography}\endgroup

\end{document}